\documentclass[10pt, draftclsnofoot, twocolumn]{IEEEtran}
\pdfoutput=1
\usepackage{amssymb}
\usepackage{algorithm}
\usepackage{algpseudocode}
\usepackage{algorithmicx}
\usepackage{amsfonts}
\usepackage{amsmath}
\usepackage{amssymb}
\usepackage{amsthm}
\usepackage{array}
\usepackage{bbding}
\usepackage{bigints}
\usepackage{booktabs}
\usepackage{cite}
\usepackage{cleveref}
\usepackage{geometry}
\usepackage{color}
\usepackage{diagbox}
\usepackage{epsfig,latexsym}
\usepackage{epstopdf}
\usepackage{graphicx}
\usepackage{fancyhdr}
\usepackage{float}
\usepackage{indentfirst}
\usepackage{lastpage}
\usepackage{makecell}
\usepackage{mathtools}
\usepackage{multirow}
\usepackage{pifont}
\usepackage{psfrag}
\usepackage{setspace}
\usepackage{stfloats}
\renewcommand{\baselinestretch}{1}
\usepackage{empheq}
\usepackage{enumerate}
\usepackage{times}
\usepackage{subfigure}
\newcolumntype{C}{>{\centering\arraybackslash$}p{\linewidth}<{$}}

\newtheorem{theorem}{Theorem}

\newtheorem{lemma}{Lemma}

\newtheorem{corollary}{Corollary}

\DeclareMathOperator{\diag}{diag}
\DeclareMathOperator{\tr}{tr}

\newcommand{\RNum}[1]{\uppercase\expandafter{\romannumeral #1\relax}}

\newcounter{eqncnt}

\newcounter{eqnback}
\begin{document}

\title{Performance Analysis and Optimization of STAR-RIS-Aided Cell-Free Massive MIMO Systems Relying on Imperfect Hardware}

\author{{Zeping Sui}, {\em Member,~IEEE}, {Hien Quoc Ngo}, {\em Fellow,~IEEE}, {Michail Matthaiou}, {\em Fellow,~IEEE}, and {Lajos Hanzo}, {\em Life Fellow,~IEEE}
\vspace{-3em}
\thanks{Zeping Sui was with the Centre for Wireless Innovation (CWI), Queen's University Belfast, Belfast BT3 9DT, UK. He is now with the School of Computer Science and Electronics Engineering, University of Essex, Colchester CO4 3SQ, UK. (e-mail: zepingsui@outlook.com)}
\thanks{Hien Quoc Ngo and Michail Matthaiou are with the Centre for Wireless Innovation (CWI), Queen's University Belfast, Belfast BT3 9DT, UK, and with the Department of Electronic Engineering, Kyung Hee University, Yongin-si, Gyeonggi-do 17104, South Korea. (e-mail: \{hien.ngo, m.matthaiou\}@qub.ac.uk).}
\thanks{Lajos Hanzo is with the Department of Electronics and Computer Science, University of Southampton, Southampton SO17 1BJ, UK. (e-mail: lh@ecs.soton.ac.uk).}%
\thanks{This work was supported in part by the Project REASON, an U.K. Government Funded Project under the Future Open Networks Research Challenge (FONRC) Sponsored by the Department of Science Innovation and Technology (DSIT), and in part by the U.K. Engineering and Physical Sciences Research Council (EPSRC) under Grant EP/X04047X/1. The work of Michail Matthaiou was supported by the European Research Council (ERC) under the European Union's Horizon 2020 Research and Innovation Programme under Grant 101001331. The work of Hien Quoc Ngo was supported by the U.K. Research and Innovation Future Leaders Fellowships under Grant MR/X010635/1, and a research grant from the Department for the Economy Northern Ireland under the US-Ireland R\&D Partnership Programme. L. Hanzo would like to acknowledge the financial support of the Engineering and Physical Sciences Research Council (EPSRC) projects under grant EP/Y037243/1, EP/W016605/1, EP/X01228X/1, EP/Y026721/1, EP/W032635/1, EP/Y037243/1 and EP/X04047X/1 as well as of the European Research Council's Advanced Fellow Grant QuantCom (Grant No. 789028).}
\thanks{Parts of this paper were presented at IEEE GLOBECOM 2024 \cite{sui2024star}.}
}
\maketitle

\begin{abstract}
\textcolor{black}{Simultaneously} transmitting and reflecting reconfigurable intelligent surface (STAR-RIS)-aided cell-free massive multiple-input multiple-output (CF-mMIMO) systems are investigated under spatially correlated fading channels using realistic imperfect hardware. Specifically, the transceiver distortions, \textcolor{black}{time-varying phase noise, and RIS phase shift errors} are considered. Upon considering imperfect hardware and pilot contamination, we derive a linear minimum mean-square error (MMSE) criterion-based cascaded channel estimator. Moreover, a closed-form expression of the downlink ergodic spectral efficiency (SE) is derived based on maximum ratio (MR) based transmit precoding and channel statistics, where both a finite number of access points (APs) and STAR-RIS elements as well as imperfect hardware are considered. Furthermore, by exploiting the ergodic signal-to-interference-plus-noise ratios (SINRs) among user equipment (UE), a max-min fairness problem is formulated for the joint optimization of the passive transmitting and reflecting beamforming (BF) at the STAR-RIS as well as of the power control coefficients. An alternating optimization (AO) algorithm is proposed for solving the resultant problems, where iterative adaptive particle swarm optimization (APSO) and bisection methods are proposed for circumventing the non-convexity of the RIS passive BF and the quasi-concave power control sub-problems, respectively. Our simulation results illustrate that the STAR-RIS-aided CF-mMIMO system attains higher SE than its RIS-aided counterpart. The performance of different hardware parameters is also evaluated. Additionally, it is demonstrated that the SE of the worst UE can be significantly improved by exploiting the proposed AO-based algorithm compared to conventional solutions associated with random passive BF and equal-power scenarios.
\end{abstract}
\vspace{-5mm}
\IEEEpeerreviewmaketitle

\section{Introduction}\label{Section 1}
Cell-free massive multiple-input multiple-output (CF-mMIMO) solutions constitute a compelling candidate for 6G wireless networks, since there are no cell boundaries in CF-mMIMO systems, and the inter-cell interference of conventional cellular networks can be mitigated \cite{7827017,8097026}. In CF-mMIMO systems, a large number of geographically distributed access points (APs) connect to central processing units (CPUs) via fronthaul links, and a small number of user equipments (UEs) are served in a coverage area within the same time and frequency resources \cite{zhang2020prospective,10522673}. Compared to conventional cellular mMIMO networks, CF-mMIMO systems achieve improved connectivity, enhanced energy efficiency, and uniformly good quality of service (QoS) of UEs \cite{ozdogan2019performance}. However, the spectral efficiency (SE) and macro-diversity gains of CF-mMIMO systems may suffer significantly under harsh propagation scenarios associated with long transmission distances, high path loss and spatially correlated channels \cite{9743355}.

As a promising technology, reconfigurable intelligent surfaces (RISs) can strike electromagnetic-level transmit waveform shaping without sophisticated digital signal processing schemes and extra power amplifiers \cite{huang2019reconfigurable}. Explicitly, effective passive beamforming (BF) can be achieved by adjusting the phase shifts of the impinging signals based on refractive RIS elements \cite{8811733}. Therefore, RISs can be deployed to enhance the SE, coverage and energy efficiency (EE) in the face of hostile channels \cite{10075372}. Moreover, RIS-aided systems can attain comparable performance at reduced transmit power and number of antennas compared to systems dispensing with a RIS \cite{di2020smart}. However, a conventional RIS can only reflect signals and support UEs in a half-plane at the same side of RISs \cite{9437234}. To circumvent this, simultaneously transmitting and reflecting RISs (STAR-RISs) have been proposed, where the impinging signals can be transmitted and reflected simultaneously toward UEs located at both sides of STAR-RISs, yielding higher SE than the conventional RIS-based systems \cite{9570143}. Hence, a STAR-RIS can, in principle, support full-space $360^\circ$ coverage upon adjusting the amplitudes and phases of incident signals.

Due to the above-mentioned benefits, RISs have been applied to CF-mMIMO systems, yielding improved SE under harsh propagation conditions \cite{9459505,van2021reconfigurable,10640072}. Specifically, a joint precoding framework was conceived for RIS-aided CF networks in \cite{9459505}. Later in \cite{van2021reconfigurable}, the uplink and downlink SEs of single-antenna AP-based RIS-aided CF-mMIMO (RIS-CF-mMIMO) systems were investigated while exploiting maximum ratio (MR) processing. In \cite{10167480}, the uplink SE of RIS-CF-mMIMO systems operating in the face of electromagnetic interference (EMI) was studied. Shi \emph{et al.} \cite{9779130} quantified the uplink SE of RIS-CF-mMIMO systems under Rician fading channels, where single-antenna APs and UEs were assumed. In \cite{9899454}, upon considering multiple RIS-assisted CF networks, a low-complexity two-timescale transmission protocol was proposed, where the joint optimization of BF at APs and RISs was investigated in the context of a max weighted sum-rate (WSR) problem. Then, closed-form expressions of uplink and downlink SEs of hybrid relay-reflecting intelligent surface-assisted CF-mMIMO systems were derived in \cite{9940169}, where MR processing techniques were harnessed. By investigating the effect of channel aging, a closed-form SE expression of RIS-CF-mMIMO systems was derived in \cite{10468556}. More recently, to utilize the full-space coverage of STAR-RIS, the authors of \cite{10297571} carried out a SE analysis of STAR-RIS-aided CF-mMIMO (STAR-RIS-CF-mMIMO) relying on perfect hardware. In \cite{10316600}, the joint optimization of the active BF at APs and passive BF at STAR-RISs was conceived for maximizing the WSR of STAR-RIS-CF-mMIMO systems. 

But again, the above treatises are developed based on perfect hardware, while some are conceived under the assumption of perfect channel state information (CSI). However, realistic imperfect hardware and channel estimation errors may impose severe performance loss \cite{bjornson2015massive}. Practical non-ideal hardware imposes hardware impairments (HWIs), such as oscillator phase noise, in-phase/quadrature (I/Q) imbalance and power amplifier non-linearity \cite{7421941,10577084,10163977}. Specifically, the impact of HWIs can be characterized by the error vector magnitude (EVM) model, where the transmit/received signals are multiplied by hardware quality factors and extra hardware distortion terms that are uncorrelated with the transmit/received signals \cite{papazafeiropoulos2021intelligent}. The HWIs have been investigated in the CF-mMIMO literature of \cite{fang2021cell,8891922,8734757}. Then, the downlink and uplink SE of HWI-constrained RIS-CF-mMIMO systems were analyzed under uncorrelated channels and ideal RISs in \cite{10225319,10197459}. Moreover, for RIS-aided systems, the RIS phase error introduced by the quantized phase drifts also results in significant performance erosion \cite{papazafeiropoulos2021intelligent}. However, to the best of our knowledge, no authors have investigated the system performance of STAR-RIS-CF-mMIMO systems under both realistic HWIs, RIS phase errors, and imperfect CSI.

To address the aforementioned issues, the contributions of our paper are  boldly and explicitly contrasted to the state-of-the-art in Table \ref{table1}, which are also detailed as follows:
\begin{table*}[t]
\footnotesize
\centering
\caption{Contrasting contributions to the existing literature related to RIS-CF networks.}
\label{table1}
\begin{tabular}{l|c|c|c|c|c|c|c|c|c|c|c|c}
\hline
Contributions & \textbf{This paper} & \cite{9459505} & \cite{van2021reconfigurable} & \cite{10167480} & \cite{9779130} & \cite{9899454} & \cite{9940169} & \cite{10468556} & \cite{10297571} & \cite{10316600} & \cite{10225319} & \cite{10197459} \\
\hline
\hline
STAR-RIS-CF-mMIMO & \checkmark & & & & & & & &\checkmark & \checkmark & \\  
\hline
Transceiver HWIs & \checkmark & & & & & & & & & & \checkmark & \checkmark \\  
\hline
RIS phase shift errors & \checkmark & & & & & & & & & & \\  
\hline
Spatially correlated channels & \checkmark &  & \checkmark & \checkmark & \checkmark & \checkmark & \checkmark & \checkmark & \checkmark &  & & \\  
\hline
Imperfect CSI & \checkmark & \checkmark & \checkmark & \checkmark & \checkmark &  & \checkmark & \checkmark & \checkmark & & \checkmark & \checkmark \\  
\hline
Closed-form expression of SE & \checkmark &  & \checkmark & \checkmark & \checkmark & \checkmark & \checkmark &  \checkmark & \checkmark &  & \checkmark & \checkmark \\  
\hline 
STAR-RIS passive BF design & \checkmark &  & & & &  & & & \checkmark  & \checkmark & & \\  
\hline
Power control design & \checkmark &  &  & \checkmark & & \checkmark & & \checkmark &  & & \\ 
\hline
Joint design for optimization & \checkmark & \checkmark &  &  & & \checkmark & & & & \checkmark & & \\ 
\hline
Alternative optimization (AO) & \checkmark & \checkmark &  &  & & \checkmark & & &  & \checkmark & \\ 
\hline
Complexity analysis of AO & \checkmark & \checkmark &  & & & \checkmark & &  & & \checkmark &   \\
\hline
Convergence analysis of AO & \checkmark & \checkmark &  & & & \checkmark & &  & & \checkmark &   \\
\hline
\end{tabular}
\vspace{-1em}
\end{table*}
\begin{itemize}
	\item We first develop the linear minimum mean-square error (MMSE) estimate of the STAR-RIS-aided cascaded channel upon considering transceiver HWIs, \textcolor{black}{nonlinear distortions}, RIS phase error and spatially correlated channel, where multi-antenna APs and pilot contamination from arbitrary pilot reuse patterns are considered. Consequently, sufficient statistical information can be provided for the ensuing data processing.
	\item We derive a closed-form expression for the downlink ergodic SE of the STAR-RIS-CF-mMIMO system based on large-scale statistics, MR precoding, realistic hardware and imperfect CSI, where finite numbers of APs, UEs and STAR-RIS elements are considered. Then, we investigate the impacts of the STAR-RIS element size, \textcolor{black}{number of RIS elements,} hardware-related parameters, array gain, and channel estimation errors. 
	\item Simulation results reveal that the proposed STAR-RIS-CF-mMIMO system can significantly improve the SE of its conventional RIS-CF-mMIMO and CF-mMIMO counterparts. Moreover, the accuracy of the derived closed-form expression of the downlink SE is validated through Monte-Carlo simulations, while our STAR-RIS-CF-mMIMO systems can still attain higher SEs under the worst-case RIS phase error compared to conventional CF-mMIMO systems. Furthermore, it can be demonstrated that the UE hardware quality factor inflicts a more significant negative impact on the SE than the AP hardware quality factor. In addition, the effects of pilot contamination, the number of APs, UEs, and antennas per AP, and the RIS element size are also characterized.
	\item We then formulate a joint STAR-RIS passive BF and power control design for our signal-to-interference-plus noise ratio (SINR) max-min problem. In particular, we decompose the NP-hard SINR max-min problem into two sub-problems. Specifically, our proposed adaptive particle swarm optimization (APSO) algorithm and the bisection method are leveraged to address the resultant non-convex and quasi-concave sub-problems. Numerical results demonstrate that our AO-based algorithm can significantly enhance the max-min SE of the conventional random passive having equal-power coefficients. \textcolor{black}{Moreover, the AO algorithm with the proposed APSO algorithm is capable of achieving better convergence compared to conventional ones, since better search capability can be stroked by the proposed APSO.}
\end{itemize}
	
The rest of our paper is organized as follows: The system model is described in Section \ref{Section 2}, while Section \ref{Section 3} details the channel estimation. Then, we provide our the downlink ergodic SE analysis in Section \ref{Section 4}. Section \ref{Section 5} develops the joint optimization problem of power control and STAR-RIS passive BF for our STAR-RIS-CF-mMIMO systems. Simulation results are provided in Section \ref{Section 6}. Finally, we conclude the paper in Section \ref{Section 7}.

\emph{Notation:} The argument, expectation and trace operators are denoted by $\arg(\cdot)$, $\mathbb{E}\{\cdot\}$ and $\text{tr}(\cdot)$, respectively; $\mathcal{CN}(\pmb{a},\pmb{B})$ stands for the complex Gaussian distribution having a mean vector $\pmb{a}$ and covariance matrix $\pmb{B}$. Moreover, the transpose, conjugate transpose and magnitude operators are denoted by $(\cdot)^T$, $(\cdot)^H$ and $\left|\cdot\right|^2$, respectively; $\pmb{I}_N$ is the $N$-dimensional identity matrix, and $I_m(\vartheta)$ denotes the $m$-order Bessel function of the first kind of $\vartheta$. The Hadamard product and Kronecker product operators are denoted as $\odot$ and $\otimes$, respectively, while $a^{(l)}$ and $A^{(m,n)}$ denote the $l$th element of vector $\pmb{a}$ and the $(m,n)$th element of matrix $\pmb{A}$, respectively. Furthermore, $\left \| \cdot \right \|$ is the Euclidean norm operator. A diagonal matrix obtained from the vector $\pmb{a}$ and from the main diagonal vector of matrix $\pmb{A}$ are $\diag\{\pmb{a}\}$ and $\diag\{\pmb{A}\}$, respectively. Furthermore, $\text{vec}(\pmb{A})$ denotes the vector obtained from stacking the columns of $\pmb{A}$. The uniform distribution within a range of $[a,b]$ is denoted by $x\sim\mathcal{U}[a,b]$.
\section{System Model}\label{Section 2}
\subsection{STAR-RIS Signal Model}\label{Section 2-1}
\begin{figure}[t]
\centering
\includegraphics[width=0.8\linewidth]{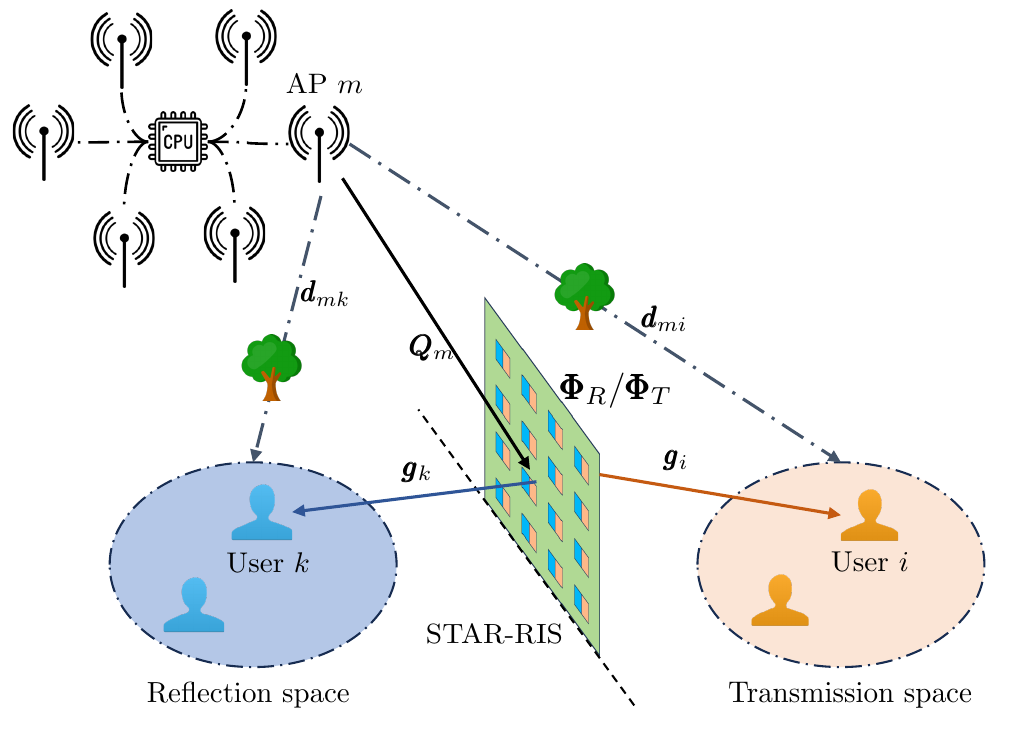}
\caption{Illustration of the considered STAR-RIS-CF-mMIMO system.}
\label{Figure1}
\end{figure}

Let us consider a STAR-RIS-CF-mMIMO system consisting of a single STAR-RIS with $N$ elements, $M$ APs equipped with $L$ antennas and $K$ UEs, where we leverage the time-division duplexing (TDD) protocol. As shown in Fig. \ref{Figure1}, the STAR-RIS divides the full plane into transmission and reflection spaces. We use $\mathcal{M}=\{1,\ldots,M\}$ to denote the AP set, while the set of STAR-RIS elements and UEs are $\mathcal{N}=\{1,\ldots,N\}$ and $\mathcal{K}=\{1,\ldots,K\}$, respectively. According to the locations of UEs and the STAR-RIS, we split the UEs into the reflection group \textcolor{black}{$\mathcal{K}_R=\{1,\ldots,K_R\}$ and the transmission group $\mathcal{K}_T=\{1,\ldots,K_T\}$}, yielding $K=K_T+K_R$. \textcolor{black}{All the APs are randomly distributed in the reflection space and are connected to the CPU via fronthaul links in the spirit of \cite{10297571}} \footnote{\textcolor{black}{Employing APs on both sides of STAR-RIS-CF-mMIMO systems is an interesting topic, which we will investigate in our future work.}}. Furthermore, the direct AP-UE links are blocked by obstacles. Let $\pmb{x}_m\in\mathbb{C}^{L}$ denote the transmit signal of the $m$th AP. The STAR-RIS reflected and transmitted signals corresponding to the $n$th STAR-RIS element is given by $\sqrt{\beta_{T,n}}e^{j{\theta}_{T,n}}\pmb{x}_m$ and \textcolor{black}{$\sqrt{{\beta}_{R,n}}e^{j{\theta}_{R,n}}\pmb{x}_m$}, respectively, where ${\beta}_{T,n}$, ${\beta}_{R,n}$ and ${\theta}_{T,n}, {\theta}_{R,n}\in[0,2\pi)$ denote the amplitude and phase shift of the $n$th STAR-RIS element responsible for the transmission and reflection spaces, respectively. The STAR-RIS passive BF matrices of the transmission and reflection spaces can be respectively formulated as $\pmb{\Phi}_T=\text{diag}\{\pmb{v}_T\}$ and $\pmb{\Phi}_R=\text{diag}\{\pmb{v}_R\}$ with $\pmb{v}_T=\left[\sqrt{\beta_{T,1}}e^{j\theta_{T,1}},\ldots,\sqrt{\beta_{T,N}}e^{j\theta_{T,N}}\right]$ and $\pmb{v}_R=\left[\sqrt{\beta_{R,1}}e^{j\theta_{R,1}},\ldots,\sqrt{\beta_{R,N}}e^{j\theta_{R,N}}\right]$. We harness the energy splitting (ES) STAR-RIS protocol, and all the STAR-RIS elements are assumed to opreate in reflection and transmission modes, simultaneously. Based on $\beta_{T,n}$ and $\beta_{R,n}$, the energy of the STAR-RIS incident waves is partitioned into transmitted and reflected components. Moreover, we assume that ${\theta}_{T,n}$ and ${\theta}_{R,n}$ are independent, while the amplitudes $\beta_{T,n}$ and $\beta_{R,n}$ are coupled according to the law of energy conservation, which satisfy $\beta_{T,n},\beta_{R,n}\in[0,1]$, $\theta_{T,n}, \theta_{R,n}\in[0,2\pi)$, $\beta_{T,n}+\beta_{R,n}=1, \forall n\in\mathcal{N}$ \cite{9570143}.	
\subsection{Phase Shift Error Model}\label{Section 2-2}
We consider realistic scenarios in which the precisions of RIS phase shifts are finite, yielding phase errors \cite{papazafeiropoulos2021intelligent}. Specifically, the phase errors can be modeled as $\tilde{\pmb{\Phi}}_k=\diag\left(e^{j\tilde{\theta}_{k,1}},\ldots,e^{j\tilde{\theta}_{k,N}}\right)$, where $\tilde{\theta}_{k,n}$ denotes the phase error of the $n$th RIS element with $k\in\{T,R\}$. The probability density function (PDF) of $\tilde{\theta}_{k,n}$ is symmetric with a zero-mean direction, and we have $\arg\left(\mathbb{E}\left\{e^{j\tilde{\theta}_{k,n}}\right\}\right)=0$. The phase errors $\tilde{\theta}_{k,n}\in[-\pi,\pi)$ are identically and independently distributed (i.i.d.) random variables (RVs) that follow either the Von Mises distribution or uniform distribution \cite{papazafeiropoulos2021intelligent}:
\begin{enumerate}[(1)]
\item Von Mises distribution: the phase errors have a zero mean with the concentration parameter $\vartheta$, representing the phase estimation accuracy. The characteristic function (CF) of phase errors is given by $\varsigma_p=I_1\left(\vartheta\right)/I_0\left(\vartheta\right)$, with $\vartheta=1/\sigma^2_p$ and the RIS phase noise power $\sigma^2_p$.
\item uniform distribution: random phase errors without \emph{a priori} information with $\vartheta=0$.
\end{enumerate}
\subsection{Channel Model}\label{Section 2-3}
Let us consider a block fading channel associated with the length of the coherence interval, i.e., the number of channel uses of $\tau_c=B_cT_c$, where $B_c$ and $T_c$ represent the coherence bandwidth and coherence time, respectively \cite{bjornson2015massive}. Moreover, the number of channel uses for the uplink channel estimation is $\tau_p$, yielding the remaining $\tau_d=\tau_c-\tau_p$ channel uses for downlink data transmission. We assume that the APs and the STAR-RIS use uniform linear arrays (ULAs) with $L$ elements and a uniform squared planar array (USPA) with $N$ elements, respectively. As shown in Fig. \ref{Figure1}, the non-line-of-sight (NLoS) direct AP-UE link can be formulated as $\pmb{d}_{mk}\sim\mathcal{CN}(\pmb{0},\pmb{R}^d_{mk})$, where $\pmb{R}^d_{mk}\in\mathbb{C}^{L\times L}$ is the spatial correlation matrix, and $\beta^d_{mk}=\tr(\pmb{R}^d_{mk})/L$ denotes the large-scale fading coefficient. Moreover, we assume that the STAR-RIS is close to the UEs, resulting in LoS channel components for the RIS-UE links. Then, the AP-RIS and RIS-UE channels can be formulated as ${\pmb{Q}}_m=\sqrt{\xi_m}\pmb{R}_{\text{A},m}^{1/2}\pmb{V}_m\pmb{R}_\text{S}^{1/2}\in\mathbb{C}^{L\times N}$ and $\pmb{g}_k=\sqrt{\frac{\alpha_k}{\iota_k+1}}\left(\sqrt{\iota_k}\bar{\pmb{g}}_k+\tilde{\pmb{g}}_k\right)\in\mathbb{C}^{N}$, respectively \cite{10297571}, where $\xi_m$ and $\alpha_k$ are the large-scale fading coefficients, while $\iota_k$ is the Rician coefficient. Moreover, $\pmb{R}_{\text{A},m}\in\mathbb{C}^{L\times L}$ and $\pmb{R}_\text{S}\in\mathbb{C}^{N\times N}$ are the deterministic Hermitian positive semi-definite spatially correlation matrices of the AP $m$ and the STAR-RIS with $\text{vec}\left(\pmb{V}_m\right)\sim\mathcal{CN}(\pmb{0},\pmb{I}_{NL})$. Note that $\pmb{R}_{\text{A},m}$ can be obtained based on \cite{hoydis2013massive}. Upon considering a multi-path transmission scenario of uniformly distributed isotropic scattering, the elements of $\pmb{R}_\text{S}$ are given by $R_\text{S}^{(a,b)}=d_H d_V\text{sinc}(2||\pmb{u}_{a}-\pmb{u}_{b}||^2/\lambda)$ with $\pmb{u}_l=[0,\mod(l-1,N_H)d_H,\lfloor(l-1)/N_H\rfloor d_V]^T$, $l\in\{a,b\}$, where $d_H$ and $d_V$ denote the horizontal width and the vertical height of each STAR-RIS element, while $N_H=\sqrt{N}$ denotes the number of RIS elements in each row \cite{bjornson2020rayleigh}. The LoS component $\bar{\pmb{g}}_k$ can be formulated based on \cite{9973349}. Furthermore, the NLoS component is given by $\tilde{\pmb{g}}_k=\pmb{R}_{\text{S}}^{1/2}\pmb{c}_k\in\mathbb{C}^{N}$ with $\pmb{c}_k\sim\mathcal{CN}(\pmb{0},\pmb{I}_N)$. Then, the $L\times 1$-dimensional cascaded channel spanning from AP $m$ to UE $k$ can be written as $\pmb{f}_{mk}=\pmb{d}_{mk}+\pmb{Q}_m\pmb{\Phi}_{k}\tilde{\pmb{\Phi}}_{k}\pmb{g}_k$, where we have $k\in\{T,R\}$ of $\pmb{\Phi}_{k}\tilde{\pmb{\Phi}}_{k}$ based on the location of UE $k$. By introducing $\pmb{G}_k=\bar{\pmb{g}}_k\bar{\pmb{g}}_k^H$, the covariance matrix of the cascaded channel $\pmb{R}_{mk}=\mathbb{E}\{\pmb{f}_{mk}\pmb{f}_{mk}^H\}$ can be formulated based on Lemma 1 \textcolor{black}{shown in Appendix A} as
\begin{align}\label{eq_R}
	{\pmb{R}}_{mk}&=\pmb{R}^d_{mk}+\pmb{R}_{\text{A},m}\tr\left(\bar{\pmb{R}}^f_{mk}\right)+\pmb{R}_{\text{A},m}\tr\left(\tilde{\pmb{R}}^f_{mk}\right)\nonumber\\
	&=\pmb{R}^d_{mk}+\pmb{R}_{\text{A},m}\tr\left({\pmb{R}}^f_{mk}\right),
\end{align}
where $\bar{\pmb{R}}^f_{mk}={\xi_m\alpha_k\iota_k}\pmb{R}_\text{S}{\pmb{\Phi}}_k\tilde{\pmb{G}}_k{\pmb{\Phi}}_k^H/\varpi_k$ and $\tilde{\pmb{R}}^f_{mk}={\xi_m\alpha_k}\pmb{R}_\text{S}{\pmb{\Phi}}_k\tilde{\pmb{R}}_\text{S}{\pmb{\Phi}}_k^H/\varpi_k$ with $\varpi_k=\iota_k+1$. Moreover, we can obtain $\tilde{\pmb{G}}_k=\mathbb{E}\left\{\tilde{\pmb{\Phi}}_k\pmb{G}_k\tilde{\pmb{\Phi}}_k^H\right\}$ and $\tilde{\pmb{R}}_\text{S}=\mathbb{E}\left\{\tilde{\pmb{\Phi}}_k{\pmb{R}}_\text{S}\tilde{\pmb{\Phi}}_k^H\right\}$. The elements of $\tilde{\pmb{G}}_k$ and $\tilde{\pmb{R}}_\text{S}$ are, respectively, given by as $\tilde{G}_{k}^{(n,n')}=G_{k}^{(n,n')}\mathbb{E}\left\{e^{j\tilde{\theta}_{k,n}-j\tilde{\theta}_{k,n'}}\right\}$ and $\tilde{R}_{\text{S}}^{(n,n')}=R_{\text{S}}^{(n,n')}\mathbb{E}\left\{e^{j\tilde{\theta}_{k,n}-j\tilde{\theta}_{k,n'}}\right\}$. Therefore, we can obtain $\tilde{\pmb{G}}_k=\varsigma^2_p{\pmb{G}}_k+(1-\varsigma^2_p)\diag({\pmb{G}}_k)$ and $\tilde{\pmb{R}}_\text{S}=\varsigma^2_p{\pmb{R}}_\text{S}+(1-\varsigma^2_p)\diag({\pmb{R}}_\text{S})$.
\subsection{Phase Noise Model}\label{Section 2-4}
Consider a realistic scenario, where both the APs and UEs are equipped with free-running oscillators, leading to transceiver phase noise terms $\phi_m^\text{AP}(t)$ and $\phi_k^\text{UE}(t)$ at the channel use $t\in\{0,\ldots,\tau_c-1\}$ that are generated during the baseband signal up-conversion and the passband signal down-conversion process. The phase noise samples can be modelled by Wiener RVs, yielding \cite{fang2021cell,bjornson2015massive}
\begin{align}\label{eq_wiener}
	\phi_m(t)&=\phi_m(t-1)+\zeta_{\phi_m}(t),\ \zeta_{\phi_m}(t)\sim\mathcal{N}(0,\varrho^2_{\phi_m}),\nonumber\\
	\psi_k(t)&=\phi_k(t-1)+\zeta_{\psi_k}(t),\ \zeta_{\psi_k}(t)\sim\mathcal{N}(0,\varrho^2_{\psi_k}),
\end{align}
where $\varrho^2_{\phi_m}$ and $\varrho^2_{\psi_k}$ are the phase noise variances corresponding to the additive terms $\zeta_{\phi_m}(t)$ and $\zeta_{\psi_k}(t)$, respectively. For ease of analysis, we stipulate $\varrho^2_{\phi_m}=\varrho^2_{\phi}$ and $\varrho^2_{\psi_k}=\varrho^2_{\psi}$, $\forall k,m$. Explicitly, the variances are given by $\varrho^2_{i}=4\pi^2 f_c^2 c_iT_s$ where $f_c$ is the carrier frequency, $c_i$ denotes a constant of the oscillator with $i\in\{\phi,\psi\}$, and $T_s$ represents the symbol duration \cite{bjornson2015massive}. Let $\varphi_{mk}(t)=\phi_m(t)+\psi_k(t)$ denote the overall transceiver oscillator phase noise, yielding $\varphi_{mk}(t)\sim\mathcal{N}[\varphi_{mk}(t-1),\delta^2]$ with $\delta^2=\varrho^2_{\phi}+\varrho^2_{\psi}$. Consequently, the effective channel at the channel use $t$ can be written as $\pmb{h}_{mk}(t)=\pmb{f}_{mk}e^{j\varphi_{mk}(t)}$.
\section{Channel Estimation}\label{Section 3}
We assume that $\tau_p<K$ pilot sequences $\{\bar{\pmb{\phi}}_u\in\mathbb{C}^{\tau_p},u=1,\ldots,\tau_p\}$ are used by the $K$ UEs in each coherence interval with $||\bar{\pmb{\phi}}_u||^2=\tau_p$, yielding pilot contamination. Let $\mathcal{P}_k\subset\{1,\ldots,K\}$ denote the UE set that utilizes the same pilot symbol as UE $k$, including the $u$th UE itself. The pilot invoked by UE $k$ is given by $\bar{\pmb{\phi}}_{u_k}$ with $u_k\in\{1,\ldots,\tau_p\}$. Hence, we have the pilot reuse pattern $\bar{\pmb{\phi}}^H_{u_k}\bar{\pmb{\phi}}_{u_i}=\tau_p,\ i\in\mathcal{P}_k$ and $\bar{\pmb{\phi}}^H_{u_k}\bar{\pmb{\phi}}_{u_i}=0,\ i\notin\mathcal{P}_k$. During the channel estimation  slot of $t\in[0,\tau_p-1]$, we assume that the phase noise remains fixed since $\tau_p\ll\tau_c$ \cite{fang2021cell}. Both transmitter and receiver distortions are considered under realistic transceiver scenarios. According to the EVM model, the received signal at the $m$th AP with $t=0$ is given by \cite{10225319,8891922}
\begin{align}\label{eq_Z}
	\pmb{Y}_m^p(0)&=\sqrt{\gamma_T}\sum_{i=1}^{K}\pmb{f}_{mi}e^{j\varphi_{mi}(0)}\bar{\pmb{x}}_i+\pmb{G}_{m}^{\text{AP}}+\pmb{N}_m,
\end{align}
where $\bar{\pmb{x}}_i=\sqrt{{p}\gamma_R}\bar{\pmb{\phi}}_{u_i}^H+\left(\pmb{w}_{i}^{\text{UE}}\right)^H$, while $\gamma_R$ and $\gamma_T\in[0,1]$ denote the hardware quality factors at the UE and the AP sides, respectively. Explicitly, the worst-case hardware is represented by $\gamma_R=\gamma_T=0$, where no information signals are received by the APs. Moreover, $\gamma_R=\gamma_T=1$ denotes the ideal hardware case, and ${p}>0$ denotes the transmitted pilot power, while the elements of the noise term $\pmb{N}_m\in\mathbb{C}^{L\times \tau_p}$ are independent and identically distributed (i.i.d.) Gaussian RVs with zero mean and a variance of $\sigma^2$. Specifically, $\pmb{w}_{i}^{\text{UE}}\sim\mathcal{CN}(\pmb{0},(1-\gamma_R){p}\pmb{I}_{\tau_p})$ is the transmitter distortion at the UE side, while the columns of the receiver distortion term $\pmb{G}_{m}^{\text{AP}}$ can be formulated as $\pmb{g}_{m}^{\text{AP}}|\{\pmb{f}_{mi}\}\sim\mathcal{CN}\left(\pmb{0},(1-\gamma_T){p}\sum_{i=1}^{K}\diag\{\pmb{f}_{mi}\pmb{f}_{mi}^H\}\right)$ \cite{10577084,10163977}. Then, we correlate the normalized pilot signal and the received signal as $\pmb{y}_{mk}^p(0)\triangleq\pmb{Y}^p_m(0)\bar{\pmb{\phi}}_{u_k}/\sqrt{\tau_p}$ to estimate the effective channel $\pmb{h}_{mk}(0)$, yielding 
\begin{align}\label{eq_z}
	\pmb{y}_{mk}^p(0)&=\sum_{i\in\mathcal{P}_k}\sqrt{\gamma_T\gamma_R{p}\tau_p}\pmb{h}_{mi}(0)+\sqrt{\frac{\gamma_T}{\tau_p}}\sum_{i=1}^K\pmb{h}_{mi}(0)\nonumber\\
	&\times\left(\pmb{w}_{i}^{\text{UE}}\right)^H\bar{\pmb{\phi}}_{u_k}+\frac{\pmb{G}_{m}^{\text{AP}}\bar{\pmb{\phi}}_{u_k}}{\sqrt{\tau_p}}+\pmb{n}_{mk},
\end{align}
where we have $\pmb{n}_{mk}\triangleq{\pmb{N}_m\bar{\pmb{\phi}}_{u_k}}/{\sqrt{\tau_p}}\sim\mathcal{CN}(\pmb{0},\sigma^2\pmb{I}_L)$, while ${\pmb{G}_{m}^{\text{AP}}\bar{\pmb{\phi}}_{u_k}}/{\sqrt{\tau_p}}|\{\pmb{f}_{mi}\}$ obeys the same distribution as $\pmb{g}_{m}^\text{AP}|\{\pmb{f}_{mi}\}$. By using the linear MMSE channel estimation technique, we can obtain the estimate of the cascaded channel as ${\pmb{h}}_{mk}(0)=\sqrt{\gamma_T\gamma_R{p}\tau_p}\pmb{R}_{mk}\pmb{\Psi}_{mk}^{-1}\pmb{z}_{mk}(0)$, where
\begin{align}\label{eq_Psi}
	\pmb{\Psi}_{mk}&=\gamma_T\gamma_R{p}\tau_p\sum_{i\in\mathcal{P}_k}\pmb{R}_{mi}+(1-\gamma_R)\gamma_T{p}\sum_{i=1}^K\pmb{R}_{mi}\nonumber\\
	&+(1-\gamma_T){p}\sum_{i=1}^K\diag(\pmb{R}_{mi})+\sigma^2\pmb{I}_L.
\end{align}
It can be readily shown that both $\pmb{\Psi}_{mk}$ and $\pmb{R}_{mk}$ are deterministic Hermitian matrices. Denoting the channel estimation error as $\tilde{\pmb{h}}_{mk}(0)=\pmb{h}_{mk}(0)-\hat{\pmb{h}}_{mk}(0)$, we have $\hat{\pmb{h}}_{mk}(0)\sim\mathcal{CN}({\pmb{0}},\pmb{\Omega}_{mk})$ and $\tilde{\pmb{h}}_{mk}(0)\sim\mathcal{CN}(\pmb{0},\pmb{C}_{mk})$ with $\pmb{\Omega}_{mk}=\gamma_T\gamma_R{p}\tau_p\pmb{R}_{mk}\pmb{\Psi}_{mk}^{-1}\pmb{R}_{mk}$ and $\pmb{C}_{mk}=\pmb{R}_{mk}-\pmb{\Omega}_{mk}$. It should be noted that $\hat{\pmb{h}}_{mk}(0)$ and $\tilde{\pmb{h}}_{mk}(0)$ are independent vectors. Moreover, we have $\mathbb{E}\left\{\hat{\pmb{h}}_{mk}^H(0)\hat{\pmb{h}}_{mk}(0)\right\}=\tr\left(\pmb{\Omega}_{mk}\right)$ and $\mathbb{E}\left\{\tilde{\pmb{h}}_{mk}^H(0)\tilde{\pmb{h}}_{mk}(0)\right\}=\tr\left(\pmb{C}_{mk}\right)$.
\section{Downlink Data Transmission and Performance Analysis}\label{Section 4}
In this section, we investigate the downlink data transmission phase in the face of imperfect hardware and derive a closed-form expression of the downlink ergodic SE using MR precoding and arbitrary STAR-RIS passive BF.
\subsection{Downlink Data Transmission}\label{Section 4-1}
During downlink data transmission $t\in[\tau_p,\tau_c-1]$, we exploit the estimated channels $\hat{\pmb{h}}_{mk}(0)$ to formulate MR precoding vectors. The downlink transmitted signal of AP $m$ can be formulated as
\begin{align}\label{eq_trans_x}
	\pmb{x}_m(t)&=\sqrt{\gamma_T\rho}\sum_{k=1}^K\sqrt{\eta_{mk}}\hat{\pmb{h}}_{mk}(0)s_k(t)+\pmb{\mu}_{m}^\text{AP}(t)\nonumber\\
	&=\bar{\pmb{x}}_m(t)+\pmb{\mu}_{m}^\text{AP}(t),
\end{align}
where $\bar{\pmb{x}}_m(t)=\sqrt{\gamma_T\rho}\sum_{k=1}^K\sqrt{\eta_{mk}}\hat{\pmb{h}}_{mk}(0)s_k(t)\in\mathbb{C}^L$, while $\rho$ is the normalized downlink SNR. Moreover, the symbol transmitted from the $k$th UE can be denoted by $s_k$ with $\mathbb{E}\{|s_k(t)|^2\}=1$, and $\eta_{mk}\geq0$ represents the corresponding power control coefficient. Given the MR precoding vector $\left\{\hat{\pmb{h}}_{mk}(0),\forall k\right\}$ and the EVM model \cite{10197459}, the transmitter hardware distortion can be formulated as $\pmb{\mu}_{m}^\text{AP}(t)\sim\mathcal{CN}\left(\pmb{0},\pmb{D}_{m,\left\{\hat{\pmb{h}}\right\}}\right)$ with the conditional correlation matrix $\pmb{D}_{m,\left\{\hat{\pmb{h}}\right\}}=(1-\gamma_T)\sum_{k=1}^K\rho\eta_{mk}\diag(\pmb{\Omega}_{mk})\in\mathbb{C}^{L\times L}$. Moreover, based on the power constraint $\mathbb{E}\{||\bar{\pmb{x}}_m(t)||^2\}\leq\gamma_T\rho$ \cite{8097026}, we have $\sum_{k=1}^K\eta_{mk}\tr(\pmb{\Omega}_{mk})\leq 1, \forall m$. The receiver distortion can be written as ${\mu}_{k}^\text{UE}(t)\sim\mathcal{CN}\left(0,(1-\gamma_R)\nu_{m,\{\hat{\pmb{h}}\}}\right)$ with 
\begin{align}\label{eq_re_distortion}
	\nu_{m,\{\hat{\pmb{h}}\}}&=\mathbb{E}\left\{\left|\sum_{m=1}^M\pmb{h}_{mk}^H(t)\pmb{x}_m(t)\right|^2|\{\pmb{h}_{mk}(t)\}\right\}\nonumber\\
	&=\mathbb{E}\left\{\sum_{m=1}^M\sum_{i=1}^K\rho\eta_{mi}\left(\gamma_T|\pmb{h}_{mk}^H(t)\hat{\pmb{h}}_{mi}(0)|^2\right.\right.\nonumber\\
	&+\left.\left.(1-\gamma_T)\left \|\hat{\pmb{h}}_{mi}(0)\odot\pmb{h}_{mk}(t)\right \|^2\right)\right\}.
\end{align}%
Then, the received signal of the $k$th UE can be expressed based on \eqref{eq_trans_x} as
\begin{align}\label{eq_received_y}
	{y}_k(t)&=\sum_{m=1}^M \sqrt{\gamma_R}\pmb{h}_{mk}^H(t)\pmb{x}_m(t)+{\mu}_{k}^\text{UE}(t)+{n}_k(t)\nonumber\\
	&={\tt DS}_k\cdot s_k(t)+{\tt BU}_k\cdot s_k(t)+\sum_{i\neq k}^K {\tt UI}_{ki}\cdot s_{i}(t)\nonumber\\
	&+{\tt HWI}_{k}+{\mu}_{k}^\text{UE}(t)+{n}_k(t),	
\end{align}
where the noise term is $n_k(t)\sim\mathcal{CN}(0,1)$. Let $C_{mk}(t)=\sum_{m=1}^M\sqrt{\eta_{mk}}\pmb{h}^H_{mk}(t)\hat{\pmb{h}}_{mk}(0)$, we have and
\begin{align}\label{eq_DS}
	{\tt DS}_k(t)&\triangleq\sqrt{\gamma_R\gamma_T\rho}\mathbb{E}\left\{C_{mk}(t)\right\},\\
	{\tt BU}_k(t)&\triangleq\sqrt{\gamma_R\gamma_T\rho}\left(C_{mk}(t)-\mathbb{E}\left\{C_{mk}(t)\right\}\right),\\
\label{eq_UI}
	{\tt UI}_{ki}(t)&\triangleq\sqrt{\gamma_R\gamma_T\rho}\sum_{m=1}^M\sqrt{\eta_{mi}}\pmb{h}^H_{mk}(t)\hat{\pmb{h}}_{mi}(0),\\
	{\tt HWI}_{k}(t)&\triangleq\sqrt{\gamma_R}\sum_{m=1}^M\pmb{h}^H_{mk}(t)\pmb{\mu}_{m}^\text{AP}(t),
\end{align}
represent the strength of the desired signal (DS), the beamforming uncertainty (BU), the UE interference (UI) caused by the $i$th UE and the interference imposed by transmitter HWIs, respectively.

\textcolor{black}{We emphasize that the HWI model invoked in this paper considers nonlinear transceiver distortions, which characterize the nonlinear power amplifiers \cite{10577084,10163977}. Furthermore, the distortion terms $\pmb{G}_{m}^{\text{AP}}$, $\pmb{\mu}_{m}^\text{AP}(t)$ are associated with the CSI and channel estimates \cite{bjornson2015massive,10577084,10163977,10197459}.}
\subsection{Downlink Ergodic SE Analysis with MR Precoding}\label{Section 4-2}
Based on the use-and-then forget capacity bounding technique of \cite{7827017}, the downlink ergodic SE (measured in bit/s/Hz) of the $k$th UE can be expressed as
\begin{align}\label{eq_DL_SE}
	R_k(\pmb{\eta}_k,\pmb{\Phi}_{k})=\frac{1}{\tau_c}\sum_{t=\tau_p}^{\tau_c-1}\log_2\left[1+{\tt SINR}_k(t)\right],
\end{align}
where ${\tt SINR}_k(t)$ denotes the effective signal-to-interference-plus-noise ratio (SINR) associated with UE $k$ at the channel use $t$, yielding
\begin{align}\label{eq_SINR}
	{\tt SINR}_k(t)=\frac{|{\tt DS}_k(t)|^2}{D_k(t)},
\end{align}
where $D_k(t)=\mathbb{E}\{|{\tt BU}_k(t)|^2\}+\sum_{i\neq k}^K\mathbb{E}\{|{\tt UI}_{ki}(t)|^2\}+\mathbb{E}\{|{\tt HWI}_k(t)|^2\}+\mathbb{E}\{|{\mu}_{k}^\text{UE}(t)|^2\}+1$.

\emph{Proposition 1:} Upon exploiting the MR precoding with arbitrary STAR-RIS passive BF $\pmb{\Phi}_k$, the closed-form expression of the downlink ergodic SE of UE $k$ can be derived in \eqref{eq_DL_SE}, where the SINR can be formulated as
\begin{align}\label{eq_SE_closed}
	{\tt SINR}_k(t)=\frac{\gamma_R\gamma_T\rho e^{-\delta^2 t}\left|\tr(\pmb{\eta}_k^{1/2}\pmb{\Omega}_k)\right|^2}{D_k(t)},
\end{align}
where $D_k(t)$ can be rewritten as \eqref{eq_Ik} shown at the top of next page, and we have $\tilde{\gamma}=(1-\gamma_R)(1-\gamma_T)$, $\pmb{\eta}_k^{1/2}=\pmb{P}_k^{1/2}\otimes\pmb{I}_L$ with $\pmb{P}_k^{1/2}=\diag\left(\sqrt{\eta_{1k}},\ldots,\sqrt{\eta_{Mk}}\right)$, $\pmb{\Omega}_k=\diag\left(\pmb{\Omega}_{1k},\ldots,\pmb{\Omega}_{Mk}\right)$. Furthermore, we have $\pmb{R}_k=\diag\left(\pmb{R}_{1k},\ldots,\pmb{R}_{Mk}\right)$, $\pmb{A}_k=\diag(a_{1k},\ldots,a_{Mk})$ with $a_{mk}=|\tr(\pmb{\Omega}_{mk})|^2$, $\pmb{B}_{ik}=\diag(b_{1ik},\ldots,b_{Mik})$ with $b_{mik}=\left|\gamma_T\gamma_R p\tau_p\tr(\pmb{R}_{mi}\pmb{\Psi}_{mk}^{-1}\pmb{R}_{mk})\right|^2$, and $\pmb{\Xi}_{ik}=\diag(\pmb{\Xi}_{1ik},\ldots,\pmb{\Xi}_{Mik})$ with $\pmb{\Xi}_{mik}=\pmb{R}_{mi}\pmb{R}_{mk}^{-1}$.
\setcounter{eqnback}{\value{equation}} \setcounter{equation}{15}
\begin{figure*}[!t]
\begin{align}\label{eq_Ik}
D_k(t)&=\gamma_T\gamma_R\rho e^{-\varrho^2_{\phi}t}\left(1-e^{-\varrho^2_{\psi}t}\right)\left|\tr(\pmb{\eta}_k^{1/2}\pmb{\Omega}_k)\right|^2+\gamma_T\rho\left(1-\gamma_Re^{-\varrho^2_{\phi}t}\right)\tr(\pmb{P}_k\pmb{A}_k)+\tilde{\gamma}\rho\tr\left(\pmb{\eta}_k(\diag(\pmb{\Omega}_{k}))^2\right)\nonumber\\
	&+\sum_{i\in\mathcal{P}_k\setminus\{k\}}\left\{\gamma_T\rho\left(1-\gamma_Re^{-\varrho^2_{\phi}t}\right)\tr(\pmb{P}_i\pmb{B}_{ik})+\gamma_T\gamma_R\rho e^{-\varrho^2_{\phi}t}\left|\tr\left(\pmb{\eta}_i^{1/2}\pmb{\Xi}_{ik}\pmb{\Omega}_k\right)\right|^2\right\}\nonumber\\
	&+\sum_{i=1}^K\rho\left[\gamma_T\tr(\pmb{\eta}_i\pmb{R}_k\pmb{\Omega}_i)+(1-\gamma_T)\tr\left(\pmb{\eta}_i\diag(\pmb{\Omega}_i)\diag(\pmb{R}_k)\right)\right]+1,
\end{align}
\hrulefill
\end{figure*}
\setcounter{eqncnt}{\value{equation}}
\setcounter{equation}{\value{eqnback}}

\emph{Proof:} See Appendix \ref{appendix2}. \hfill $\blacksquare$

\emph{Remark 1:} Based on Proposition 1, we can observe that all the components of \eqref{eq_SINR} are associated with the hardware quality factors and phase noise. The power of DS can be decreased by $\gamma_R\gamma_T\rho(1-e^{-\delta^2t})\left|\tr(\pmb{\eta}_k^{1/2}\pmb{\Omega}_k)\right|^2$ due to the transceiver phase noise. The increase of BU power can be formulated based on \eqref{eq_BU_final} as $\gamma_T\gamma_R\rho\left(1-e^{-\varrho^2_{\phi}t}\right)\left[e^{-\varrho^2_{\phi}t}\left|\tr(\pmb{\eta}_k^{1/2}\pmb{\Omega}_k)\right|^2+\tr(\pmb{P}_k\pmb{A}_k)\right]$, which is introduced by the realistic phase noise. The interference due to the pilot contamination can be formulated as $\gamma_T\rho\left(1-\gamma_Re^{-\varrho^2_{\phi}t}\right)\tr(\pmb{P}_i\pmb{B}_{ik})+\gamma_T\gamma_R\rho e^{-\varrho^2_{\phi}t}\left|\tr\left(\pmb{\eta}_i^{1/2}\pmb{\Xi}_{ik}\pmb{\Omega}_k\right)\right|^2$, while the interference boost inflicted by pilot contamination can be mitigated by $\sum_{i\in\mathcal{P}_k\setminus\{k\}}\left[\left|\tr\left(\pmb{\eta}_i^{1/2}\pmb{\Xi}_{ik}\pmb{\Omega}_k\right)\right|^2-\tr(\pmb{P}_i\pmb{B}_{ik})\right]\gamma_R\gamma_T\rho\\\left(1-e^{-\varrho^2_{\phi}t}\right)$.
\section{Joint Power Control and STAR-RIS Passive Beamforming Design}\label{Section 5}
In this section, we first formulate the max-min fairness optimization problem for the joint design of STAR-RIS passive BF and power control. Due to the non-convexity of the optimization problem, we then partition it into two sub-problems, and propose an AO-based algorithm to solve it.
\subsection{Problem Formulation}\label{Section 5-1}
We aim for maximizing the worst SINR ${\tt SINR}_k(t)$ at the $t$th channel use along all the UEs, upon designing the power control coefficients $\pmb{\eta}_k$ and the STAR-RIS passive BF matrix $\pmb{\Phi}_k$ for $k=1,\ldots,K$. Therefore, the joint optimization problem can be formulated as
\begin{spacing}{1.0}\setcounter{equation}{16}	
\begin{subequations}\label{eq_opt_p}
\begin{align}
	\mathop{\max }_{\pmb{\eta}_k,\pmb{\Phi}_k}\min_{k} \ & {\tt SINR}_k(t)\label{eq_opt_p1} \\
	{\text{s.t.}}\;\;
	& \beta_{T,n}+\beta_{R,n}=1, \forall n\in\mathcal{N},\label{eq_opt_p2}\\
	&\beta_{k,n}\in[0,1], \theta_{k,n}\in[0,2\pi), \forall n\in\mathcal{N},k, \label{eq_opt_p3}\\
	&\eta_{mk}\geq 0,\forall m,k,\label{eq_opt_p4}\\
	&\sum_{k=1}^K\eta_{mk}\tr(\pmb{\Omega}_{mk})\leq 1, \forall m,
\label{eq_opt_p5}				
\end{align}
\end{subequations}
\end{spacing}
\vspace{0.5em}\hspace{-1.5em} where \eqref{eq_opt_p2} and \eqref{eq_opt_p3} are constraints of the STAR-RIS passive BF matrices, while \eqref{eq_opt_p4} and \eqref{eq_opt_p5} denote the power control constraints. It can be observed that the optimization problem of \eqref{eq_opt_p} is non-convex, since its objective function and STAR-RIS passive BF-related constraints are all non-convex. Therefore, we solve the max-min SINR problem of \eqref{eq_opt_p} by an AO technique. The passive BF and power control coefficients are optimized alternately, while the other variables are fixed. Explicitly, the optimization problem of \eqref{eq_opt_p} can be decomposed into two sub-problems as follows.
\subsection{The Sub-Problem of STAR-RIS Passive BF}\label{Section 5-2}
When the power control coefficients are fixed, the sub-problem of STAR-RIS passive BF can be cast as
\begin{spacing}{1}
\begin{subequations}\label{eq_sub3}	
\begin{align}
	\mathop{\max }_{\beta_{T,n},\beta_{R,n},\theta_{T,n},\theta_{R,n}}\min_{k} \ & {\tt SINR}_k(t)\label{eq_sub3_p1} \\
	{\text{s.t.}}\;\;
	& \eqref{eq_opt_p2},\eqref{eq_opt_p3}\label{eq_sub3_p2}.	
\end{align}
\end{subequations}
\end{spacing}
\hspace{-1.0em}It can be readily shown that the objective function and passive BF-related constraints of \eqref{eq_sub3} are all non-convex, and the passive BF matrices $\pmb{\Phi}_k$ for $k=1,\ldots,K$ among different UEs are highly coupled in our SINR expression of \eqref{eq_SE_closed}. Hence, it is hard (if not possible) to tackle the optimization problem of \eqref{eq_sub3} upon invoking conventional optimization algorithms \textcolor{black}{and its global optimal solution of is nontrivial to attain.} Therefore, we propose an APSO algorithm to optimize the amplitudes and phase shifts of the STAR-RIS. Let us denote the passive BF parameter set as $\mathcal{X}=\{\beta_{T,n},\beta_{R,n},\theta_{T,n},\theta_{R,n}:\forall n\in\mathcal{N}\}$. Based on \eqref{eq_opt_p2} and \eqref{eq_opt_p3}, a total of $3N$ parameters have to be optimized. Let us introduce $\pmb{x}=\{\pmb{\beta}_{T},\pmb{\theta}_T,\pmb{\theta}_R\}$ with $\pmb{\beta}_{T}=\{{\beta}_{T,1},\ldots,{\beta}_{T,N}\}$, $\pmb{\theta}_{T}=\{{\theta}_{T,1},\ldots,{\theta}_{T,N}\}$ and $\pmb{\theta}_{R}=\{{\theta}_{R,1},\ldots,{\theta}_{R,N}\}$. Let us furthermore consider the PSO population set $\mathcal{L}$ with $L_P$ particles, and exploit $\beta_{T,n}\sim\mathcal{U}[0,1]$ and $\theta_{k,n}\sim\mathcal{U}[0,2\pi)$, $\forall n\in\mathcal{N},k\in\{T,R\}$ at the initial iteration of the APSO. During the $t_P$th APSO iteration, the $l_P$th particle can be formulated as $\pmb{x}_{l_P}^{(t_P)}=\left\{\pmb{\beta}_{T,l_P}^{(t_P)},\pmb{\theta}_{T,l_P}^{(t_P)},\pmb{\theta}_{R,l_P}^{(t_P)}\right\}=\left\{{x}_{l_P,1}^{(t_P)},\ldots,{x}_{l_P,3N}^{(t_P)}\right\}$ for $l_P=1,\ldots,L_P$, and all the $L_P$ particles are updated at the same time. Based on the SINR values that the particles attain, the APSO algorithm records both the global and local optimal particles, yielding $\pmb{x}^{(t_P)}_\text{gbest}$ and $\pmb{x}^{(t_P)}_{l_P,\text{pbest}}$, respectively. The velocities of particles can be updated as $\pmb{v}^{(t_P +1)}_{l_P}=\omega^{(t_P)}\pmb{v}^{(t_P)}_{l_P}+\pmb{\Delta}_{l_P}^{(t_P)}$, where $\pmb{\Delta}_{l_P}^{(t_P)}=c_1r_1\left(\pmb{x}^{(t_P)}_{l_P,\text{pbest}}-\pmb{x}^{(t_P)}_{l_P}\right)+c_2r_2\left(\pmb{x}^{(t_P)}_{\text{gbest}}-\pmb{x}^{(t_P)}_{l_P}\right)$ with the acceleration coefficients $c_1=c_2=1.496$ \cite{985692}, $r_a\sim\mathcal{U}[0,1]$ where $a\in\{1,2\}$. The adaptive inertia weight $\omega^{(t_P)}$ can be formulated as $\omega^{(t_P)}=\frac{e^{6\kappa_t}-1}{e^{6\kappa_t}+1}(\omega_\text{max}-\omega_\text{min})+\omega_\text{min}$, where $\kappa_t=(T_P-t_P)/{T_P}$ with the maximum number of iterations $T_P$, while $\omega_\text{min}$ and $\omega_\text{max}$ denote the initial and final values of the inertia weight, respectively. Then the particles can be updated as $\pmb{x}_{l_P}^{(t_P+1)}=\pmb{x}_{l_P}^{(t_P)}+\pmb{v}^{(t_P +1)}_{l_P}$. After $T_P$ iterations, the globally optimal particle $\pmb{x}^{(T_P)}_\text{gbest}$ is selected to formulate the STAR-RIS passive BF $\pmb{\Phi}_T$ and $\pmb{\Phi}_R$. Our proposed APSO procedure is summarized in Algorithm \ref{alg1}.
\begin{algorithm}[htbp]
\footnotesize
\caption{APSO algorithm for optimizing STAR-RIS passive BF in the $t_A$th AO iteration}
\label{alg1}
\begin{algorithmic}[1]
    \Require The obtained power control efficient $\pmb{z}^{[t_A-1]}_k$, $c_1$, $c_2$, $\omega_{\text{min}}$ and $\omega_{\text{max}}$.      
    \State \textbf{Preparation}: Set the maximum number of iterations $T_{P}$.
    \State \textbf{Initialize} The passive BF $\pmb{x}_{l_p}^{(0)}$ for $l_P=1,\ldots,L_P$.
    \For{$t_P=0$ to $T_{P}$}
    \For{$l_P=1$ to $L_P$}
    \State Calculate ${\tt SINR}_k(t)^{(t_P)}$ based on particle $\pmb{x}_{l_p}^{(t_P)}$.
    \EndFor
    \State Obtain $\pmb{x}_{l_P,\text{pbest}}^{(t_P)}$ and $\pmb{x}_{\text{gbest}}^{(t_P)}$.
    \State $\omega^{(t_P)}=\frac{e^{6\kappa_t}-1}{e^{6\kappa_t}+1}(\omega_\text{max}-\omega_\text{min})+\omega_\text{min}$.
    \For{$l_P=1$ to $L_P$}
    \State Calculate $\pmb{\Delta}_{l_P}^{t_P}$.
    \State $\pmb{v}^{(t_P +1)}_{l_P}=\omega^{(t_P)}\pmb{v}^{(t_P)}_{l_P}+\pmb{\Delta}_{l_P}^{(t_P)}$.
    \State Updated particles $\pmb{x}_{l_P}^{(t_P+1)}=\pmb{x}_{l_P}^{(t_P)}+\pmb{v}^{(t_P +1)}_{l_P}$.
    \EndFor
\EndFor
\State \textbf{return} passive BF $\pmb{\Phi}_T^{[t_A]}$ and $\pmb{\Phi}_R^{[t_A]}$ based on $\pmb{x}^{(T_P)}_\text{gbest}$.
\end{algorithmic}
\end{algorithm}
\subsection{The Sub-Problem of Power Control}\label{Section 5-3}
Fix the STAR-RIS passive BF coefficients and let $z_{mk}\triangleq\sqrt{\eta_{mk}}$, then we can formulate the sub-problem of power control by introducing the slack variables $\varepsilon_{mk}$, $\varkappa_{ik}$, $q_{mk}$ and $r_{mk}$, yielding
\begin{spacing}{0.6}
	\begin{subequations}\label{eq_sub2}	
\begin{align}
	\mathop{\max }_{\mathcal{X}_{mik}}\min_{k} \ & \overline{\tt {SINR}}_k(t)\label{eq_sub2_p1} \\
	{\text{s.t.}}\;\;
	&z_{mk}\geq 0,\forall m,k,\label{eq_sub2_p2}\\
	&\sum_{k=1}^Kz^2_{mk}\tr(\pmb{\Omega}_{mk})\leq 1, \forall m,\label{eq_sub2_p3}\\	
	&\sum_{i\in\mathcal{P}_k\setminus\{k\}}z^2_{mi}b_{mik}\leq \varepsilon^2_{mk},\forall m,k,\label{eq_sub2_p4}\\
	&\sum_{m=1}^M z_{mi}\tr(\pmb{\Xi}_{mik}\pmb{\Omega}_{mk})\leq \varkappa_{ik},\forall i\neq k,\label{eq_sub2_p5}\\	
	&\sum_{i=1}^K z_{mi}^2\tr(\pmb{\Gamma}_{mk})\leq q^2_{mk},\forall m,k,\label{eq_sub2_p6}\\		
	&\sum_{i=1}^K z_{mi}^2\tr(\pmb{R}_{mk}\pmb{\Omega}_{mi})\leq r^2_{mk},\forall m,k,\label{eq_sub2_p7}	
\end{align}
\end{subequations}
\end{spacing}
\hspace{-1.0em}where $\mathcal{X}_{mik}=\{z_{mk},\varepsilon_{mk},\varkappa_{ik},q_{mk},r_{mk}\}$, $\pmb{\Gamma}_{mk}=\diag(\pmb{\Omega}_{mi})\diag(\pmb{R}_{mk})$ and we have $\overline{\tt {SINR}}_k(t)={{\gamma_R\gamma_T\rho e^{-\delta^2 t}}\left[\sum_{m=1}^M{z_{mk}}\tr(\pmb{\Omega}_{mk})\right]^2}/{\bar{D}_k(t)}$ with $\bar{D}_k(t)$ of \eqref{eq_barD} shown at the top of next page.
\setcounter{eqnback}{\value{equation}} \setcounter{equation}{19}
\begin{figure*}[!t]
\begin{align}\label{eq_barD}
	\bar{D}_k(t)&=\gamma_R\gamma_T\rho e^{-\varrho^2_{\phi}t}\left(1-e^{-\varrho^2_{\psi}t}\right)\left[\sum_{m=1}^Mz_{mk}\tr(\pmb{\Omega}_{mk})\right]^2+\gamma_T\rho\left(1-\gamma_Re^{-\varrho^2_{\phi}t}\right)\sum_{m=1}^M{z_{mk}^2}\left|\tr(\pmb{\Omega}_{mk})\right|^2\nonumber\\
	&+\tilde{\gamma}\rho\sum_{m=1}^M{z_{mk}^2}\tr\left(\diag(\pmb{\Omega}_{mk})^2\right)+\gamma_T\rho\left(1-\gamma_Re^{-\varrho^2_{\phi}t}\right)\sum_{m=1}^M	\varepsilon_{mk}^2+\gamma_T\gamma_R\rho e^{-\varrho^2_{\phi}t}\sum_{i\in\mathcal{P}_k\setminus\{k\}}\varkappa_{ik}^2\nonumber\\
	&+(1-\gamma_T)\rho\sum_{m=1}^Mq^2_{mk}+\gamma_T\rho\sum_{m=1}^Mr^2_{mk}+1.
		\end{align}
\hrulefill
\end{figure*}
\setcounter{eqncnt}{\value{equation}}
\setcounter{equation}{\value{eqnback}}

The problem of \eqref{eq_sub2} can be reformulated as
\begin{spacing}{0.6}\setcounter{equation}{20}
	\begin{subequations}\label{eq_sub4}	
\begin{align}
	\mathop{\max }_{\mathcal{X}_{mik},u}\ & u\label{eq_sub4_p1} \\
	{\text{s.t.}}\;\;
	&u\leq\overline{\tt {SINR}}_k(t)\label{eq_sub4_p2}, k=1,\ldots,K,\\
	&\eqref{eq_sub2_p2}-\eqref{eq_sub2_p7}.\label{eq_su4_p3}	
\end{align}
\end{subequations}
\end{spacing}

\emph{Proposition 2:} The objective function and the problem of \eqref{eq_sub4} are all quasi-concave.

\textcolor{black}{\emph{Proof:} The proof is provided in Appendix \ref{appendix3}.} \hfill $\blacksquare$

Therefore, based on the SOCP feasibility bisection search of \cite{boyd2004convex}, we can efficiently solve the problem in \eqref{eq_sub4}. Specifically, we solve a sequence of convex feasibility problems during each iteration, and the bisection method is detailed in Algorithm \ref{alg2}. \textcolor{black}{Specifically, the parameters $u_\text{max}$ and $u_\text{min}$ can be attained based on the maximum and minimum values of the worst SINR, upon invoking equal-power control (EPC) and random passive beamforming schemes.}
\begin{algorithm}[htbp]
\footnotesize
\caption{Bisection algorithm for optimizing power control coefficients in the $t_A$th AO iteration}
\label{alg2}
\begin{algorithmic}[1]
    \Require The passive BF $\pmb{\Phi}_T^{[t_A-1]}$ and $\pmb{\Phi}_R^{[t_A-1]}$, $u_{\text{max}}$ and $u_{\text{min}}$ which denote the range of values of \eqref{eq_sub2_p1}.     
    \State \textbf{Preparation}: Set the tolerance parameter $\epsilon_\text{bi}>0$. 
    \State \textbf{Initialize} Set $u=\frac{u_{\text{min}}+u_{\text{max}}}{2}$.
    \State Solve the following convex feasibility problem:
    \begin{align}
    	    \begin{cases}
    	\left\|\pmb{v}_k^{[t_A]}\right\|\leq\sqrt{\tilde{u}(t)\gamma_R\gamma_T\rho e^{-\varrho^2_{\phi}t}}\sum\limits_{m=1}^M{z_{mk}^{[t_A]}}\tr(\pmb{\Omega}_{mk}),\forall k,\\
    	z_{mk}^{[t_A]}\geq 0,\forall m,k,\\
    	\sum_{k=1}^K\left(z_{mk}^{[t_A]}\right)^2\tr(\pmb{\Omega}_{mk})\leq 1,\forall m,\\
    	\sum_{i\in\mathcal{P}_k\setminus\{k\}}\left(z_{mi}^{[t_A]}\right)^2 b_{mik}\leq \varepsilon^2_{mk},\forall m,k,\\
    	\sum_{m=1}^M z_{mi}^{[t_A]}\tr(\pmb{\Xi}_{mik}\pmb{\Omega}_{mk})\leq \varkappa_{ik},\forall i\neq k,\\
    	\sum_{i=1}^K \left(z_{mi}^{[t_A]}\right)^2\tr(\diag(\pmb{\Omega}_{mi})\diag(\pmb{R}_{mk}))\leq q^2_{mk},\forall m,k,\\
    	\sum_{i=1}^K \left(z_{mi}^{[t_A]}\right)^2\tr(\pmb{R}_{mk}\pmb{\Omega}_{mi})\leq r^2_{mk},\forall m,k,\\
    	0<u<\frac{e^{-\varrho^2_{\psi}t}}{1-e^{-\varrho^2_{\psi}t}},
    	    	    	    	    \end{cases}\nonumber
    \end{align}
    where we have $\tilde{u}(t)=\left[\left(\frac{1}{u}+1\right)e^{-\varrho_{\psi}^2 t}-1\right]$, $\pmb{v}_k\triangleq[\pmb{v}_{k1};\pmb{v}_{k2};\pmb{v}_{k3};\pmb{v}_{k4};\pmb{v}_{k5};\pmb{v}_{k6};1]$ with $\pmb{v}_{k1}=\sqrt{\gamma_T\rho\left(1-\gamma_Re^{-\varrho^2_{\phi}t}\right)}(\pmb{z}_k\odot\pmb{\lambda}_{k1})$, $\pmb{v}_{k2}=\sqrt{\tilde{\gamma}\rho}(\pmb{z}_k\odot\pmb{\lambda}_{k2})$, $\pmb{v}_{k3}=\sqrt{\gamma_T\rho}\pmb{r}_k$, $\pmb{v}_{k4}=\sqrt{\gamma_T\rho\left(1-\gamma_Re^{-\varrho^2_{\phi}t}\right)}\pmb{\varepsilon}_k$, $\pmb{v}_{k5}=\sqrt{\gamma_T\gamma_R\rho e^{-\varrho^2_{\phi}t}}\pmb{\varkappa}_{k}$ and $\pmb{v}_{k6}=\sqrt{(1-\gamma_T)\rho}\pmb{q}_{k}$. We have $\pmb{z}_k=[z_{1k},\ldots,z_{Mk}]^T$, $\pmb{\lambda}_{k1}=\left[\left|\tr(\pmb{\Omega}_{1k})\right|,\ldots,\left|\tr(\pmb{\Omega}_{Mk})\right|\right]^T$, $\pmb{\lambda}_{k2}=[\lambda_{1k2},\ldots,\lambda_{Mk2}]^T$ with $\lambda_{mk2}=\sqrt{\tr\left(\diag(\pmb{\Omega}_{mk})^2\right)}$ for $m=1,\ldots,M$, $\pmb{r}_k=[r_{1k},\ldots,r_{Mk}]$, $\pmb{\varepsilon}_k=[\varepsilon_{1k},\ldots,\varepsilon_{Mk}]^T$, $\pmb{\varkappa}_k=\left\{{\varkappa}_{ik}|{i\in\mathcal{P}_k\setminus\{k\}},\forall i\right\}$ and $\pmb{q}_k=[q_{1k},\ldots,q_{Mk}]^T$.
    \State If the problem is feasible, set $u_{\text{min}}=u$, else set $u_{\text{max}}=u$.
    \State Stop if $u_{\text{max}}-u_{\text{min}}<\epsilon_\text{bi}$, otherwise go to step 2.
\State \textbf{return} $\pmb{z}^{[t_A]}=\left\{\pmb{z}_1^{[t_A]},\ldots,\pmb{z}_K^{[t_A]}\right\}$.
\end{algorithmic}
\end{algorithm}
\subsection{Convergence Analysis}\label{Section 5-4}
Based on the above sub-problems as well as Algorithms~\ref{alg1} and \ref{alg2}, the AO-based algorithm is proposed for optimizing the power coefficients and the passive BF. The details of the proposed AO-based procedure are summarized in Algorithm \ref{alg3}, where $F^{[t_A]}$ denotes the fitness value during the $t_A$th iteration.  Upon defining $\pmb{\eta}=\{\pmb{\eta}_1,\ldots,\pmb{\eta}_K\}$ and $\pmb{\Phi}=\{\pmb{\Phi}_1,\ldots,\pmb{\Phi}_K\}$, we can formulate $F(\pmb{\eta}^{[t_A]},\pmb{\Phi}^{[t_A]})\overset{(a)}\leq F(\pmb{\eta}^{[t_A]},\pmb{\Phi}^{[t_A+1]})\overset{(b)}\leq F(\pmb{\eta}^{[t_A+1]},\pmb{\Phi}^{[t_A+1]})$, where $(a)$ and $(b)$ hold, since the updates of $\pmb{\Phi}$ and $\pmb{\eta}$ can attain the max-min SINR, when the other variable remains fixed, respectively \footnote{\textcolor{black}{Please refer to \cite{985692} for more details on the convergence analysis of PSO.}}. Therefore, the value of the objective function in \eqref{eq_opt_p1} is monotonically nondecreasing during each AO iteration, and the convergence of our AO-based algorithm can be guaranteed.
\begin{algorithm}[htbp]
\footnotesize
\caption{AO-based algorithm for joint power control and STAR-RIS passive BF optimization}
\label{alg3}
\begin{algorithmic}[1]
    \Require Channel estimation-related matrices $\pmb{R}_{mk}$, $\pmb{\Omega}_{mk}$ and $\pmb{\Psi}_{mk}$, the parameters $t$, $\rho$, $\gamma_T$, $\gamma_R$, $\delta^2$, $\varrho_{\psi}^2$ and $\varrho_{\phi}^2$ .   
    \State \textbf{Preparation}: Set the tolerance parameter $\epsilon_\text{AO}$.
    \State \textbf{Initialize} The power control coefficients $\pmb{z}^{[0]}$, the passive BF $\pmb{\Phi}_T^{[0]}$ and $\pmb{\Phi}_R^{[0]}$.
	\State \textbf{Repeat}
	\State Update $\pmb{\Phi}_T^{[t_A]}$ and $\pmb{\Phi}_R^{[t_A]}$ upon solving \eqref{eq_sub3} based on Algorithm \ref{alg1} with $\pmb{z}^{[t_A-1]}$.
	\State Update $\pmb{z}^{[t_A]}$ by solving \eqref{eq_sub2} with Algorithm \ref{alg2} and $\pmb{\Phi}_T^{[t_A-1]}$ and $\pmb{\Phi}_R^{[t_A-1]}$.
	\State \textbf{return} Optimized power control coefficients $\pmb{z}$, passive BF $\pmb{\Phi}_T$ and $\pmb{\Phi}_R$.
	\State Compute the objective value $F^{[t_A]}$.
	\State \textbf{Until} $\left|F^{[t_A]}-F^{[t_A-1]}\right|\leq\epsilon_\text{AO}$,
	\State \textbf{return} optimized $\pmb{z}$, $\pmb{\Phi}_T$ and $\pmb{\Phi}_R$.
\end{algorithmic}
\end{algorithm}
\subsection{Complexity Analysis}\label{Section 5-5}
Let us now consider the complexity of the passive BF sub-problem of \eqref{eq_sub3}. The complexity of particle initialization is on the order of $\mathcal{O}(3NL_P)$. During each APSO iteration, sorting the fitness values has a complexity order of $\mathcal{O}(L_p\log L_P)$. The complexity of calculating the velocities is on the order of $\mathcal{O}(3NL_P)$. The complexity of calculating the inertial weight factor is given by $\mathcal{O}(T_P)$. Therefore, the overall complexity of the passive BF sub-problem is on the order of $\mathcal{O}_1=\mathcal{O}(3NL_P+(L_p\log L_P+3NL_P+1)T_P)=\mathcal{O}(T_PL_P\log L_P+3T_PNL_P)$.

Each iteration of Algorithm \ref{alg2} involves $A_v\triangleq4MK+P_kK$ real-valued scalar variables along with $P_k\triangleq|\mathcal{P}_k\setminus\{k\}|$, $A_l\triangleq P_kK$ linear constraints and $A_q\triangleq4MK+M$ quadratic constraints. The number of iterations of Algorithm \ref{alg2} can be formulated as $\log_2(\frac{u_\text{max}-u_\text{min}}{\epsilon_\text{bi}})$. Therefore, the total complexity of Algorithm \ref{alg2} is on the order of $\mathcal{O}_2=\mathcal{O}(A_v^2(A_l+A_q))\log_2(\frac{u_\text{max}-u_\text{min}}{\epsilon_\text{bi}})$ \cite{boyd2004convex}. Consequently, the total complexity of the proposed AO-based algorithm is on the order of $\mathcal{O}[T_{A}(\mathcal{O}_1+\mathcal{O}_2)]$, where $T_{A}$ denotes the number of iterations in Algorithm \ref{alg3}.

\section{Numerical Results}\label{Section 6}
\subsection{Simulation Setup}\label{Section 6-1}
In this section, numerical results are provided to characterize the overall performance of our proposed STAR-RIS-CF-mMIMO system relying on imperfect hardware. Let us consider a geographic area of $1\times 1$ $\text{km}^2$, where the locations of APs/UEs/STAR-RIS are provided in terms of $(x,y)$ coordinates in meter units. The STAR-RIS is located at the coordinate of $(x_\text{RIS},y_\text{RIS})=(0,0)$, while the reflection and transmission spaces are respectively denoted as $x<0$ and $x>0$, $\forall y$. We consider a harsh propagation scenario, where the $M$ APs are uniformly distributed in a square region of $x_{\text{AP}}\in[-500,-250]$ m and $y_{\text{AP}}\in[250,500]$ m. Moreover, the $K_R$ UEs are uniformly distributed within a sub-region of the reflection space yielding $x_R,y_R\in[-325,-125]$ m, while the remaining $K_T$ UEs are positioned within the sub-region as $x_T\in[125,325]$ m and $y_T\in[-125,-325]$ m. \textcolor{black}{Unless stated otherwise, other simulation parameters are summarized in Table \ref{table2}.} For the conventional EPC scheme, we assume that the power control coefficient of AP $m$ remains constant when associated with different UEs, i.e., we have $\eta_{mk}=\left(\sum_{i=1}^K\tr(\pmb{\Omega}_{mi})\right)^{-1},\forall m,k$. The random BF (RBF) associated with the EPC technique is utilized, when we evaluate our SE performance analysis results. The other channel-related parameters are the same as those in \cite{ozdogan2019performance}. Furthermore, we exploit the CF-mMIMO (denoted as ``Cell-free'') and the conventional RIS-CF-mMIMO (denoted as ``RIS-cell-free'') with ideal hardware as benchmarks. Explicitly, for the RIS-CF-mMIMO system, we harness two reflection-only RISs having $N/2$ elements adjacent to each other at the coordinate of $(0,0)$.
\begin{table}[htbp]
\vspace{-5mm}
\footnotesize
\centering
\caption{\textcolor{black}{Simulation parameters}}
\label{table2}
\begin{tabular}{l|l}
\hline
\textbf{Parameters} & \textbf{Values} \\
\hline
\hline
Carrier frequency, $f_c$ & 2 GHz \\  
\hline
Bandwidth, $B$ & 10 MHz \\
\hline
Length of coherence block, $\tau_c$ & 100\\
\hline
Coherence bandwidth, $B_c$ & 100 KHz\\
\hline
Coherence time, $T_c$ & 1 ms\\
\hline
Power of pilots, $p$ & 0.2 W\\
\hline
Maximum transmitted power of symbols, $\rho$ & 1 W\\
\hline
Symbol duration, $T_s$ & 10 $\mu s$\\
\hline
Oscillator constants, $c_i,\forall i$ & $1\times 10^{-18}$\\
\hline
Spacings of RIS elements, $d_{\text{RIS}}$ & $\lambda/4$\\
\hline
Spacings of AP elements, $d_{\text{AP}}$ & $\lambda/2$\\
\hline
Height of UEs & $1.5$ m\\
\hline
Height of RIS & $30$ m\\
\hline
Height of APs & $12.5$ m\\
\hline
Max inertia weight, $\omega_\text{max}$ & $0.9$\\
\hline
Min inertia weight, $\omega_\text{min}$ & $0.4$\\
\hline
Number of UEs in the reflection space, $K_R$ & $3$\\
\hline
Number of UEs in the reflection space, $K_T$ & $3$\\
\hline
Number of antennas with each AP, $L$ & $4$\\
\hline
Horizontal width of each RIS element, $d_H$ & $\lambda/4$\\
\hline
Vertical height of each RIS element, $d_V$ & $\lambda/4$\\
\hline
Number of pilots, $\tau_p$ & $3$\\
\hline
Tolerance parameter of bisection algorithm, $\epsilon_\text{bi}$ & $0.01$\\
\hline
Max number of APSO iterations, $T_P$ & $100$\\
\hline
Number of APSO particles, $L_P$ & $10$\\
\hline
RIS phase error parameter, $\vartheta$ & $3$\\
\hline
Number of APs, $M$, & $16$\\
\hline
Number of RIS elements, $N$, & $16$\\
\hline
\end{tabular}
\end{table}
\vspace{-2em}

\subsection{Performance of STAR-RIS-CF-mMIMO Systems}\label{Section 6-2}
\begin{figure}[htbp]
\vspace{-1em}
\centering
\begin{minipage}[htbp]{0.8\linewidth}
\centering
\includegraphics[width=\linewidth]{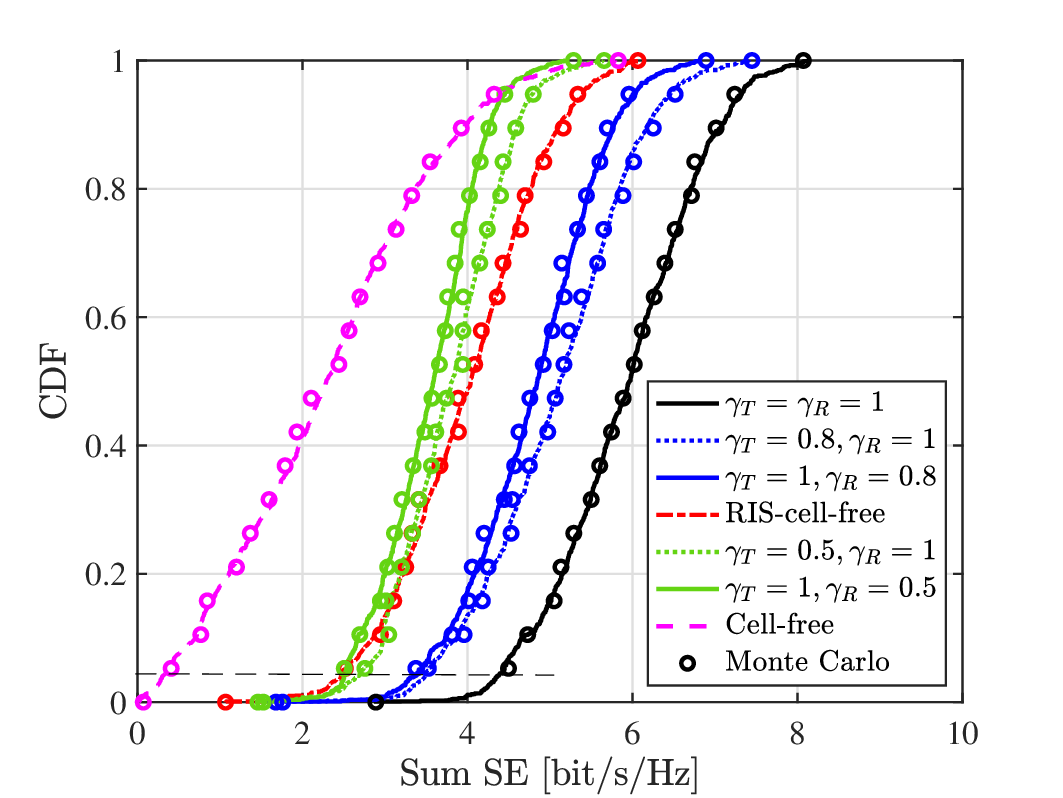}
\vspace{-2em}
\caption{CDF of the downlink sum SE for the STAR-RIS-CF-mMIMO with different $\gamma_T$ and $\gamma_R$ ($M=20,N=128,\vartheta=4$).}
\label{Figure2}
\end{minipage}
\hspace{0.01in}
\begin{minipage}[htbp]{0.8\linewidth}
\centering
\includegraphics[width=\linewidth]{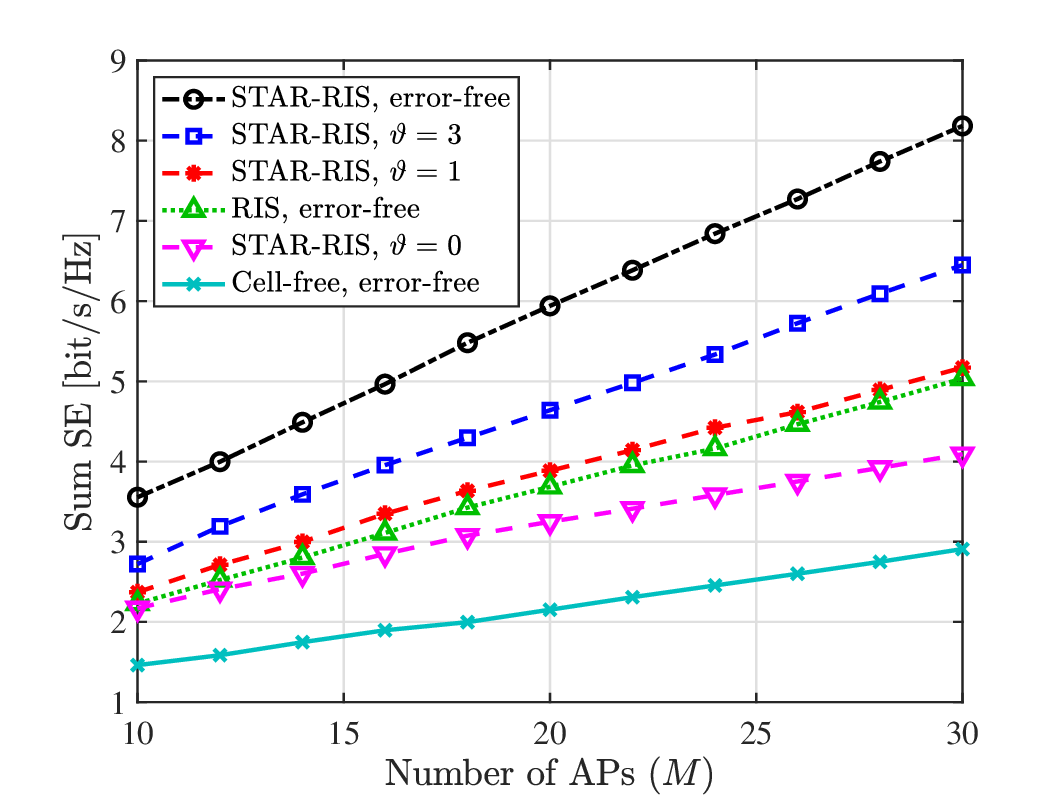}
\vspace{-2em}
\caption{\textcolor{black}{Downlink sum SE versus the number of APs $M$ operating at different RIS phase error parameter $\vartheta$ ($N=64,\gamma_T=\gamma_R=1$).}}
\label{Figure3}
\end{minipage}
\end{figure}

We investigate the cumulative distribution function (CDF) of the downlink sum SEs of our proposed STAR-RIS-CF-mMIMO systems having different hardware quality factors $\gamma_T$ and $\gamma_R$ in Fig. \ref{Figure2}. The closed-form expression of the sum SEs and the corresponding Monte-Carlo simulation results are computed based on \eqref{eq_DL_SE}, \eqref{eq_SINR} and \eqref{eq_Ik}. From Fig. \ref{Figure2}, we have the following observations. Firstly, given $\gamma_T=\gamma_R=1$, the STAR-RIS-CF-mMIMO systems are capable of attaining significant SE performance gains compared to the conventional RIS-CF-mMIMO and CF-mMIMO systems. This is because higher degrees of freedom (DoF) can be achieved by simultaneous transmission and reception using the STAR-RIS. Specifically, the $95\%$-likely SE of STAR-RIS-CF-mMIMO systems is about $14.7$ and $1.7$ times higher than that of the CF-mMIMO and RIS-CF-mMIMO systems. Moreover, the STAR-RIS-CF-mMIMO can always attain a higher sum SE than the CF-mMIMO system, even in the worst hardware case. Furthermore, the receiver hardware quality factor $\gamma_R$ may inflict a more significant negative impact on the sum SE than $\gamma_T$, implying that the UE quality is more important than its AP counterpart. This observation is consistent with the conclusions of [30]. Finally, the perfect overlap between the closed-form expression of sum SE and Monte-Carlo simulations showcases the accuracy of our analytical results, \textcolor{black}{which also implies the tightness of our derived use-and-then-forget bound.}

In Fig. \ref{Figure3}, we investigate the sum SE of the STAR-RIS-CF-mMIMO for $\gamma_T=\gamma_R=1$ versus the number of APs $M$, and we consider different values of $\vartheta$. It can be observed that higher SEs can be attained by deploying more APs, since more efficient beamforming can be achieved when we increase the number of spatial DoF. Moreover, given $M=20$, the STAR-RIS-CF-mMIMO systems having no phase errors are capable of providing about $172.7\%$ and $62.2\%$ higher sum SE compared to the CF-mMIMO and RIS-CF-mMIMO systems, respectively. Furthermore, the higher the values of $\vartheta$, the higher SE values are achieved by the STAR-RIS-CF-mMIMO, since lower phase errors are imposed on the STAR-RIS. \textcolor{black}{Specifically, the $\vartheta=3$ scenario is respectively capable of achieving $24.9\%$ and $54.8\%$ higher sum SEs than the $\vartheta=1$ and $\vartheta=0$ cases when $M=30$.} We emphasize that the STAR-RIS-CF-mMIMO system can always attain a significantly higher SE than its CF-mMIMO counterpart, even under $\vartheta=0$. \textcolor{black}{Additionally, it can be seen that the average SE per UE is less than $3$ bit/s/Hz, since we consider a harsh transmission scenario of high path loss for the direct AP-UE links.}

\begin{figure}[htbp]
\vspace{-1em}
\centering
\begin{minipage}[htbp]{0.8\linewidth}
\centering
\includegraphics[width=\linewidth]{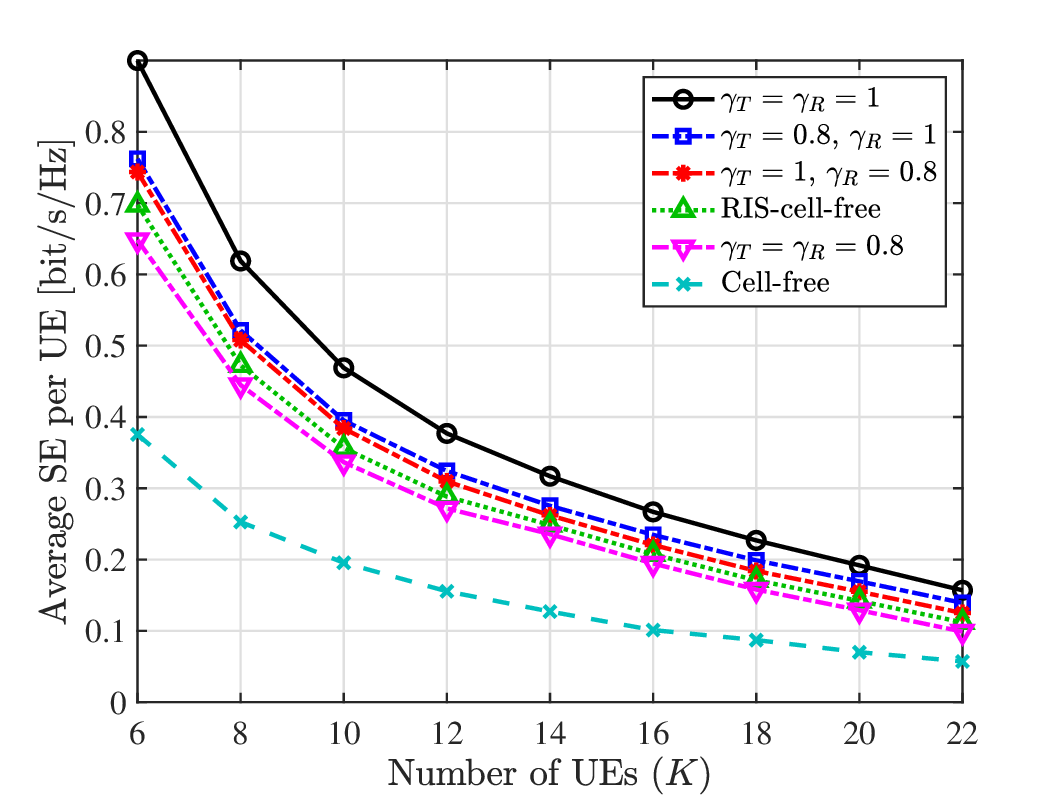}
\vspace{-2em}
\caption{\textcolor{black}{Downlink average SE per UE for STAR-RIS-CF-mMIMO, RIS-CF-mMIMO and CF-mMIMO versus the number of UEs $K$ operating at different hardware quality factors.}}
\label{Figure4}
\end{minipage}
\hspace{0.01in}
\begin{minipage}[htbp]{0.8\linewidth}
\centering
\includegraphics[width=\linewidth]{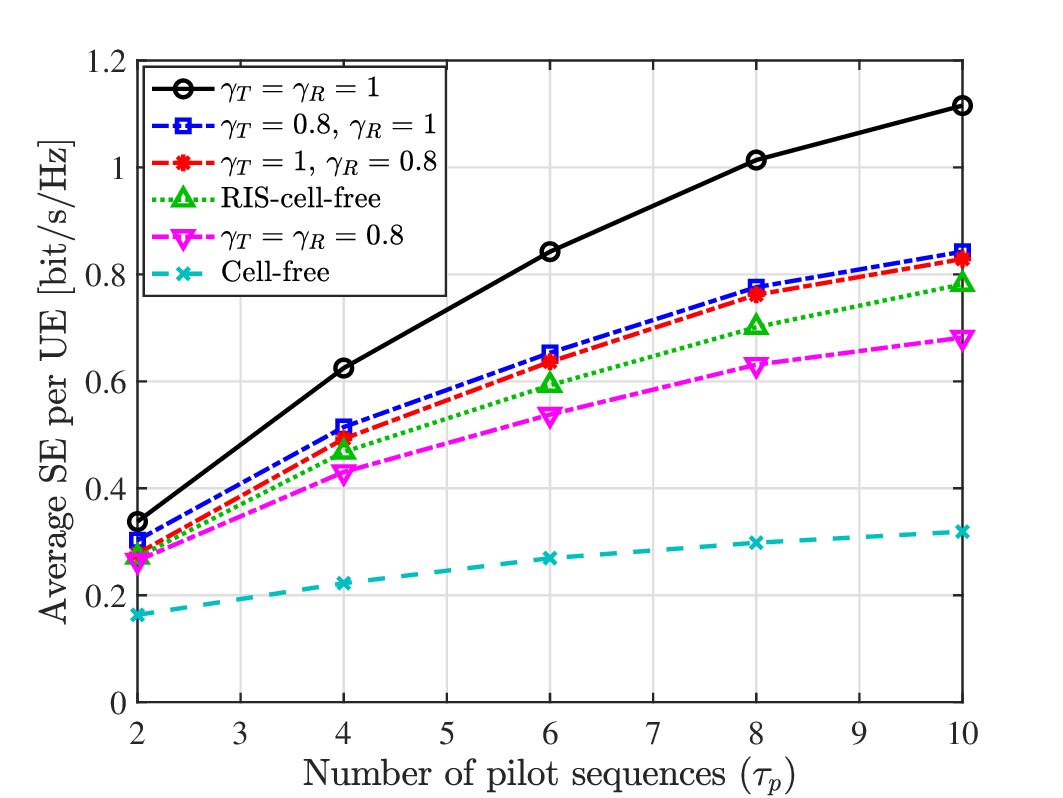}
\vspace{-2em}
\caption{Downlink average SE per UE for STAR-RIS-CF-mMIMO, RIS-CF-mMIMO and CF-mMIMO versus the number of pilot $\tau_p$ operating at different hardware quality factors with $K=10$.}
\label{Figure5}
\end{minipage}
\end{figure}

The average SE per UE is depicted as a function of the number of UEs $K$ in Fig. \ref{Figure4}, where different hardware quality factors $\gamma_T$ and $\gamma_R$ are considered. It can be observed that supporting more UEs can yield a significant SE performance erosion, since higher inter-user interference (IUI) induced pilot contamination is encountered. Moreover, the SE degradation introduced by higher IUI can be mitigated using a transceiver of better hardware quality. Furthermore, the performance gap, which is constrained by phase errors, narrows as the value of $K$ increases. \textcolor{black}{However, at a value of $K=22$, the STAR-RIS-CF-mMIMO is still capable of achieving $41.7\%$ and $183.3\%$ average SE improvements compared to its RIS-CF-mMIMO and CF-mMIMO counterparts, which underlines the importance of deploying STAR-RISs to compensate for harsh propagation.}

In Fig. \ref{Figure5}, the effect of the number of pilots $\tau_p$ is investigated for different hardware quality factors. From Fig. \ref{Figure5}, we have the following observations. Firstly, the SE of all the systems can be enhanced by using more pilots, since lower pilot contamination can be attained. Moreover, given the values of $\gamma_T$ ($\gamma_R$), lower values of $\gamma_R$ ($\gamma_T$) degrade the SE performance. This is because lower hardware quality factors introduce an avoidable transceiver HWIs. Furthermore, the SEs of STAR-RIS-CF-mMIMOs with $(\gamma_T,\gamma_R)=(0.8,1)$ and $(\gamma_T,\gamma_R)=(1,0.8)$ are still higher compared to the RIS-CF-mMIMO and CF-mMIMO systems having ideal hardware, yielding $1.03$ and $25.6$ times higher SE, respectively.
\begin{figure}[htbp]
\vspace{-1em}
\centering
\begin{minipage}[htbp]{0.8\linewidth}
\centering
\includegraphics[width=\linewidth]{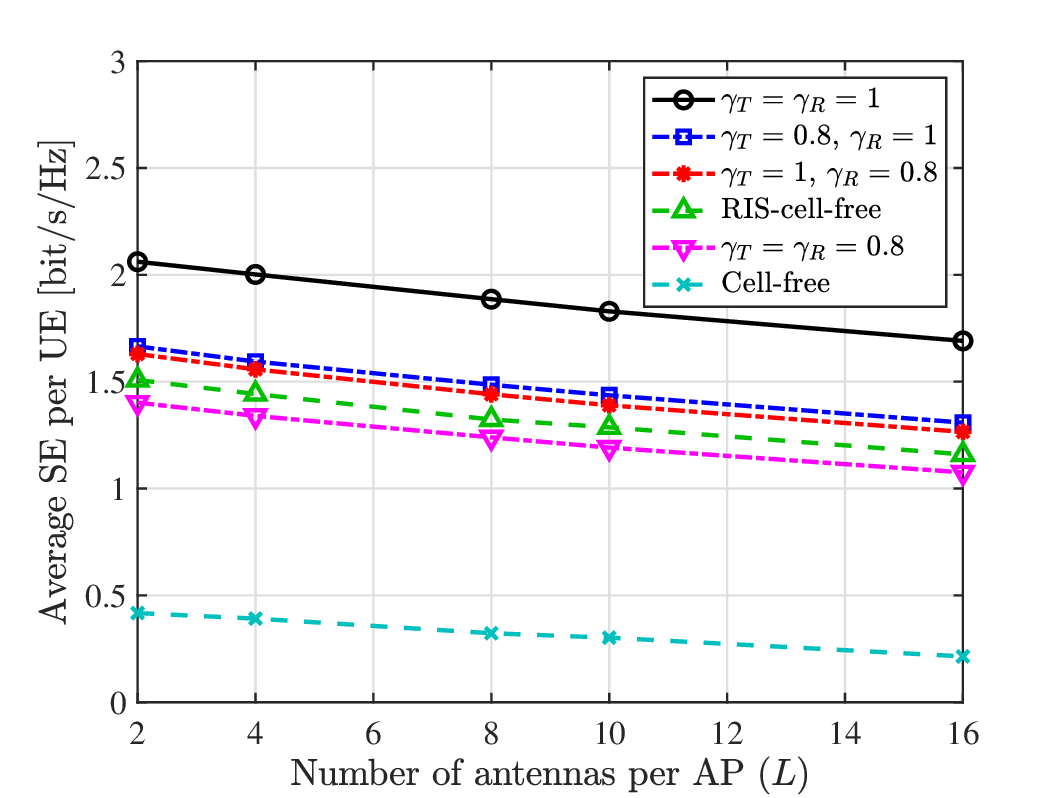}
\vspace{-2em}
\caption{Average SE per UE for STAR-RIS-CF-mMIMO, RIS-CF-mMIMO and CF-mMIMO versus the number of antennas per AP $L$ operating at different hardware quality factors and $LM=80$ ($K=4,\tau_p=4$).}
\label{Figure6}
\end{minipage}
\hspace{0.01in}
\begin{minipage}[htbp]{0.8\linewidth}
\centering
\includegraphics[width=\linewidth]{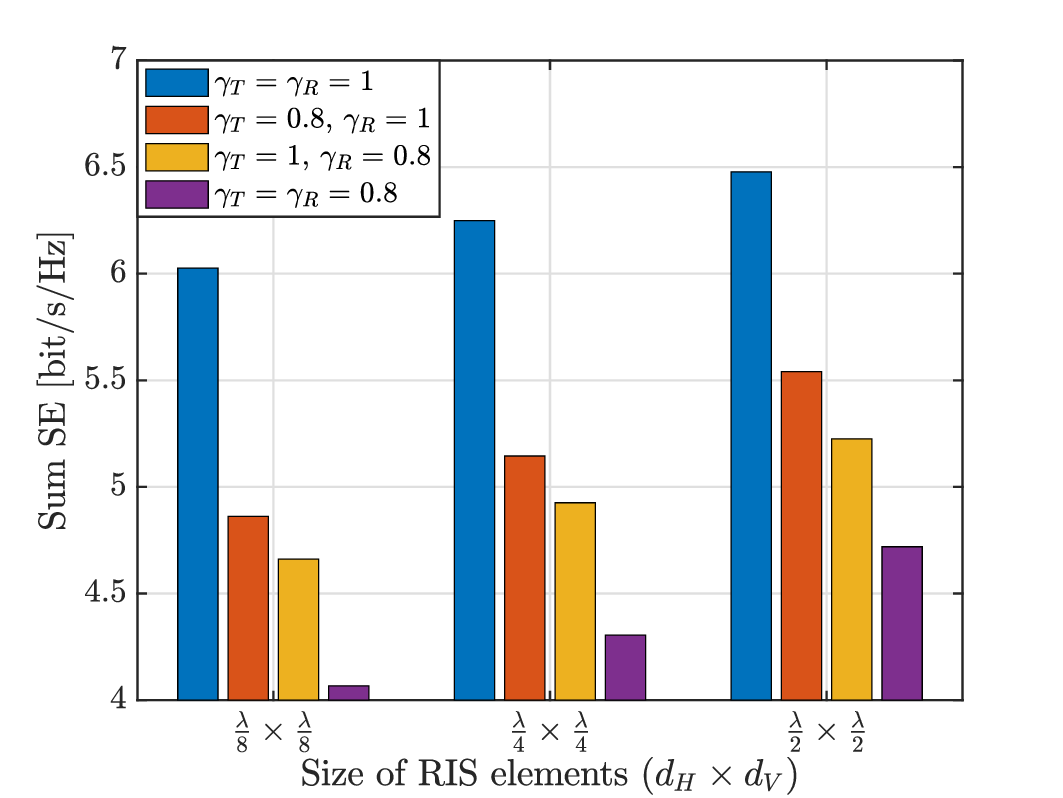}
\vspace{-2em}
\caption{Sum SE for STAR-RIS-CF-mMIMO versus the size of RIS elements $d_H\times d_V$ operating at different hardware quality factors with $K=10$.}
\label{Figure7}
\end{minipage}
\end{figure}

We compare the average SE per UE of both STAR-RIS-CF-mMIMO, RIS-CF-mMIMO and conventional CF-mMIMO systems versus different number of antennas per AP $L$ in Fig. \ref{Figure6}, while the total number of antennas remains fixed at $LM=80$. We can see from Fig. \ref{Figure6} that although the average SE of our STAR-RIS-CF-mMIMO system with $\gamma_T=\gamma_R=1$ still outperforms RIS-CF-mMIMO and CF-mMIMO, there is a significant UE erosion as $L$ increases. This observation can be explained as follows. Firstly, a higher value of $L$ implies that fewer APs are deployed, since the number of antennas is unchanged, leading to higher path losses. Secondly, a lower value of $M$ tends to yield a lower macro-diversity. Therefore, we can readily show that macro-diversity is the dominant factor for this scenario, rather than the array gain at the AP side.

\textcolor{black}{In Fig. \ref{Figure7},} we compare the downlink sum SE for different size of elements, while also different hardware quality factors are considered, and the total number of STAR-RIS elements $N$ remains unchanged. Moreover, higher values of $d_H d_V$ imply that the size of the RIS $N d_H d_V$ increases. It can be seen from Fig. \ref{Figure7} that a better sum SE performance can be attained upon increasing the size of each RIS element. Moreover, lower values of $\gamma_R$ always result in significant SE erosion, consistent with the observations shown in Fig. \ref{Figure2}-Fig. \ref{Figure6}.

\begin{figure}[htbp]
\vspace{-1.5em}
\centering
\includegraphics[width=0.8\linewidth]{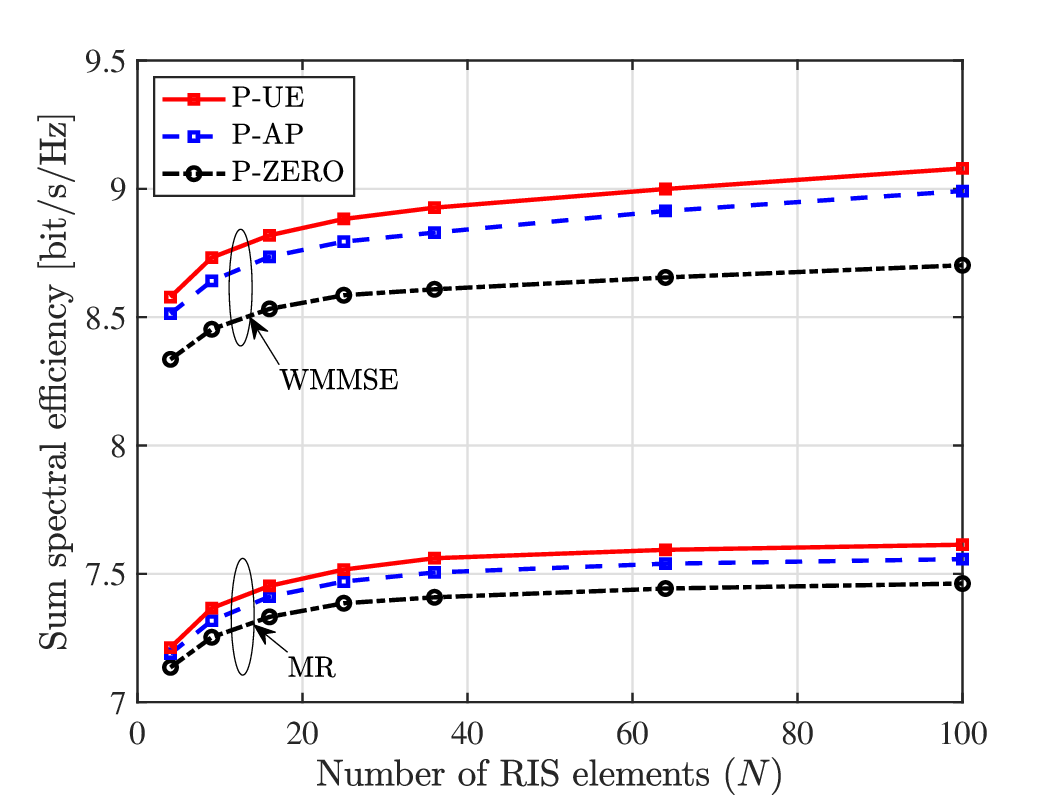}
\caption{\textcolor{black}{Sum SE with WMMSE and MR precoding against the number RIS elements $N$ over different RIS positions ($K=20,M=30,\gamma_T=\gamma_R=1,\vartheta=1$).}}
\label{Figure12}
\end{figure}

\textcolor{black}{In Fig. \ref{Figure12}, the sum SE against the number of RIS elements $N$ over different RIS deployment positions is shown, where the weighted MMSE (WMMSE) precoding \cite{10437404} is also invoked as a benchmark to further characterize the performance of our proposed STAR-RIS-CF-mMIMO. Specifically, we utilize the following three different RIS deployment positions: 1) P-UE: the RIS is deployed near the UEs, i.e., $(x_\text{RIS},y_\text{RIS})=(0,-125)$ m; 2) P-AP: the RIS is deployed near the APs, i.e., $(x_\text{RIS},y_\text{RIS})=(0,250)$ m; 3) P-ZERO: the RIS is deployed at the zero point, i.e., $(x_\text{RIS},y_\text{RIS})=(0,0)$ m. We can make the following observations from Fig. \ref{Figure12}. Firstly, given the precoding scheme and the number of RIS elements $N$, the P-UE RIS deployment can attain the highest sum SE, followed by the P-AP and P-ZERO topologies. This trend implies that the SE of our STAR-RIS-CF-mMIMO system can be enhanced by optimizing the STAR-RIS deployment location. Moreover, when the RIS deployment position and the number of RIS elements $N$ are fixed, it can be observed that the WMMSE precoding is capable of attaining higher sum SE compared to the conventional MR precoding. This is because the WMMSE technique can converge to a stationary point of the max weighted sum rate optimization problem \cite{10437404,10013728}. It should be observed that the SE gain is attained by WMMSE precoding at the cost of high complexity, which is introduced by matrix inversion \cite{10437404}. Under the MR scenario, the sum SE converges, when $N\geq36$. However, the sum SE can still be enhanced using more RIS elements in the WMMSE counterparts. This implies that the WMMSE precoding can utilize the spatial DoF provided by the STAR-RIS in the proposed system.}

\subsection{Performance of AO Algorithm}\label{Section 6-3}
\begin{figure}[htbp]
\vspace{-1em}
\centering
\begin{minipage}[htbp]{0.8\linewidth}
\centering
\includegraphics[width=\linewidth]{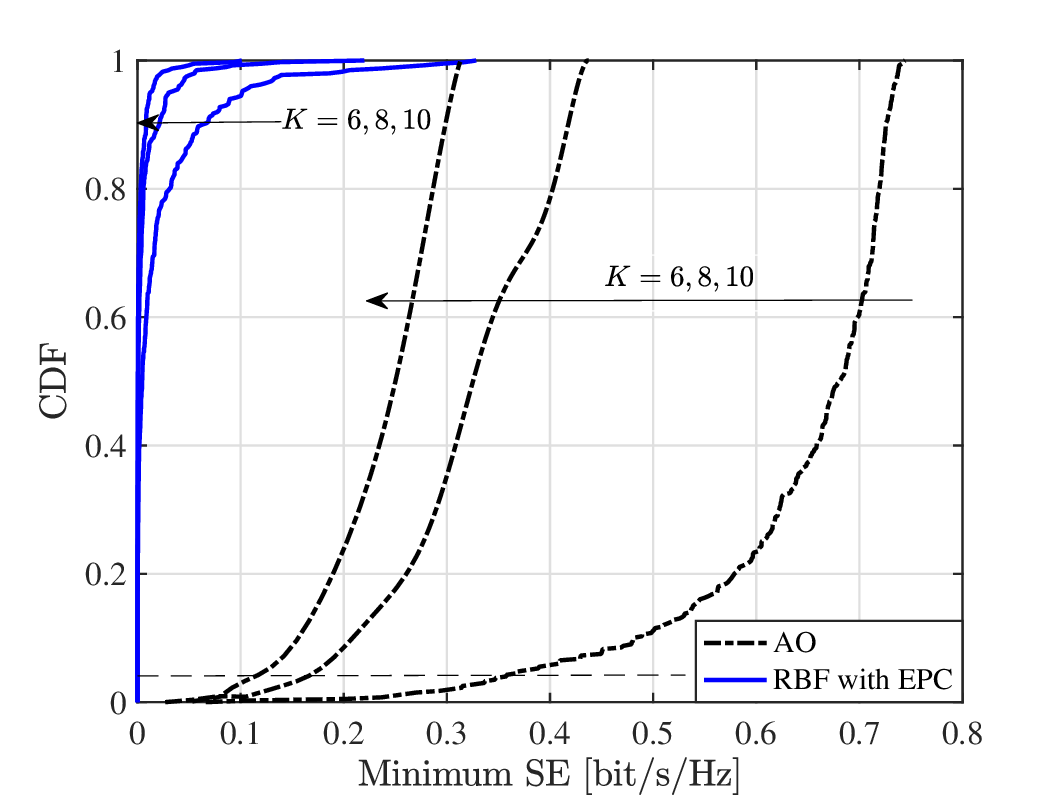}
\caption{CDF of the downlink minimum SE for the STAR-RIS-CF-mMIMO with different $K$ ($M=15,\gamma_T=\gamma_R=0.8$).}
\label{Figure8}
\end{minipage}
\hspace{0.01in}
\begin{minipage}[htbp]{0.8\linewidth}
\centering
\includegraphics[width=\linewidth]{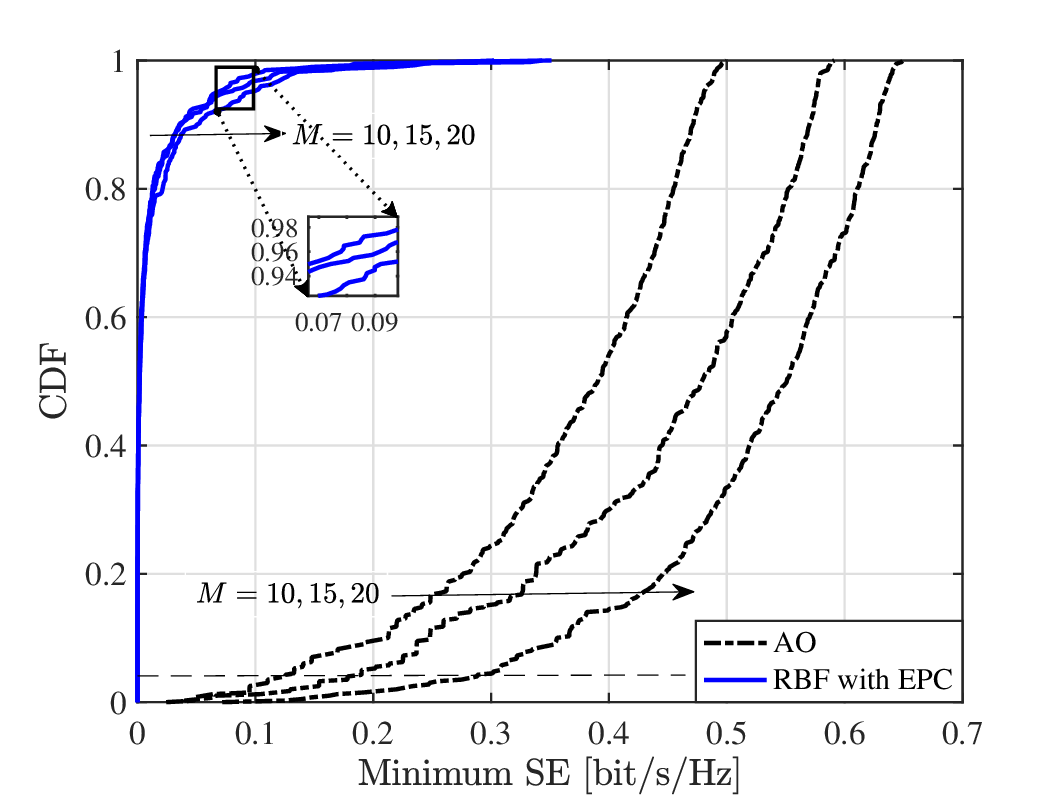}
\vspace{-2em}
\caption{CDF of the downlink minimum SE for the STAR-RIS-CF-mMIMO with different $M$ ($\gamma_T=\gamma_R=0.8$).}
\label{Figure9}
\end{minipage}
\end{figure}

The CDFs of the minimum SE using the proposed AO-based algorithm and the conventional RBF with EPC scheme for different numbers of UEs are shown in Fig. \ref{Figure8}. It can be observed that our proposed AO-based algorithm significant increases the minimum SE compared to the conventional RBF using the EPC scheme, which implies that the AO-based algorithm is capable of improving the max-min SINR fairness. Moreover, similar to Fig. \ref{Figure4}, an increase in the value of $K$ can lead to performance erosion, since higher pilot contamination and IUI must be tolerated.

The CDF of minimum SE using the AO-based algorithm and the conventional RBF with EPC are investigated \textcolor{black}{in Fig. \ref{Figure9},} where different numbers of APs $M$ are considered. It can be observed from Fig. \ref{Figure9}, that the worst SE per UE is significantly improved upon invoking the proposed AO-based algorithm, regardless of the number of APs. Moreover, increasing the number of APs is capable of achieving a higher $95\%$-likely minimum SE, which is consistent with the observation of Fig. \ref{Figure3}.
\begin{figure}[htbp]
\vspace{-1.5em}
\centering
\includegraphics[width=0.8\linewidth]{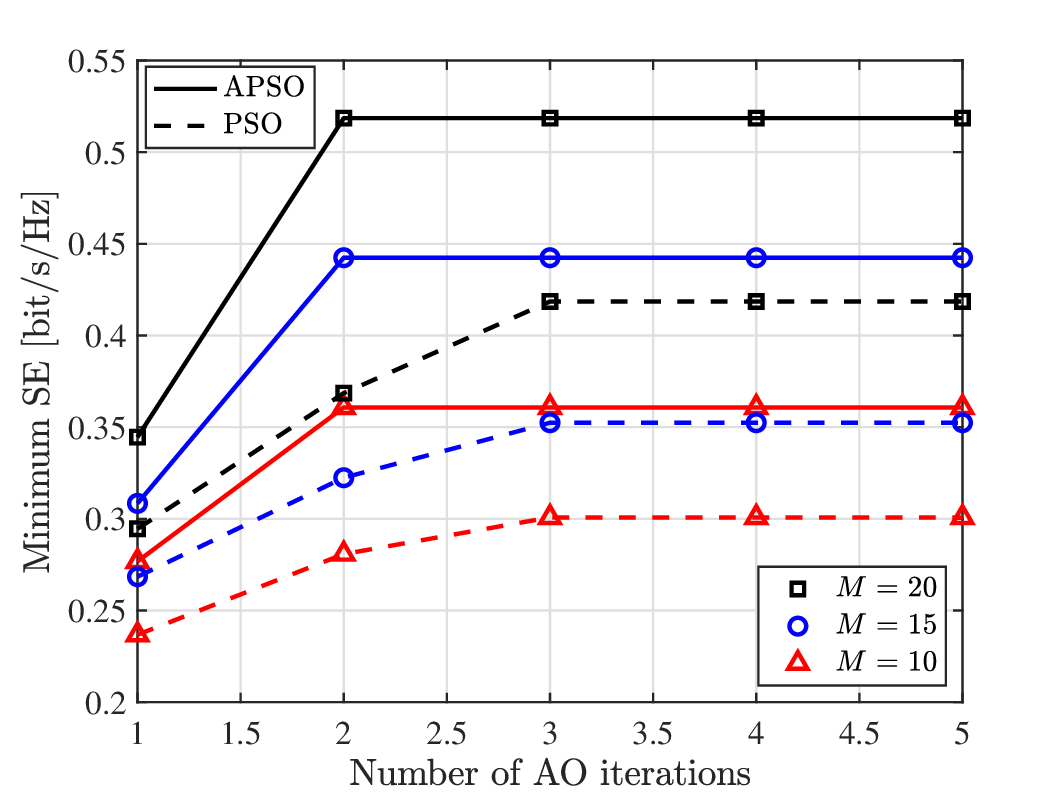}
\vspace{-1em}
\caption{\textcolor{black}{Convergence of the proposed AO algorithms using APSO and PSO for STAR-RIS-CF-mMIMO with different $M$ ($\gamma_T=\gamma_R=0.8$).}}
\label{Figure10}
\end{figure}

In Fig. \ref{Figure10}, the convergence of the proposed AO-based algorithm is plotted. Explicitly, upon solving the sub-problems alternatively, we can obtain the final solution of the joint design optimization problem. Moreover, the solution obtained from each sub-problem is invoked as the input of the other sub-problem. It is demonstrated that our AO-based algorithm using the proposed APSO can achieve convergence using as few as two AO iterations. \textcolor{black}{By contrast, the AO algorithm relying on the conventional PSO can only converge after three AO iterations. Furthermore, given the value of $M$, the minimum SE attained by the APSO-based AO algorithm is higher than its PSO-based counterpart. This is because the APSO algorithm achieves better search capability with adaptive inertia weight.}

\begin{figure}[htbp]
\centering
\includegraphics[width=0.8\linewidth]{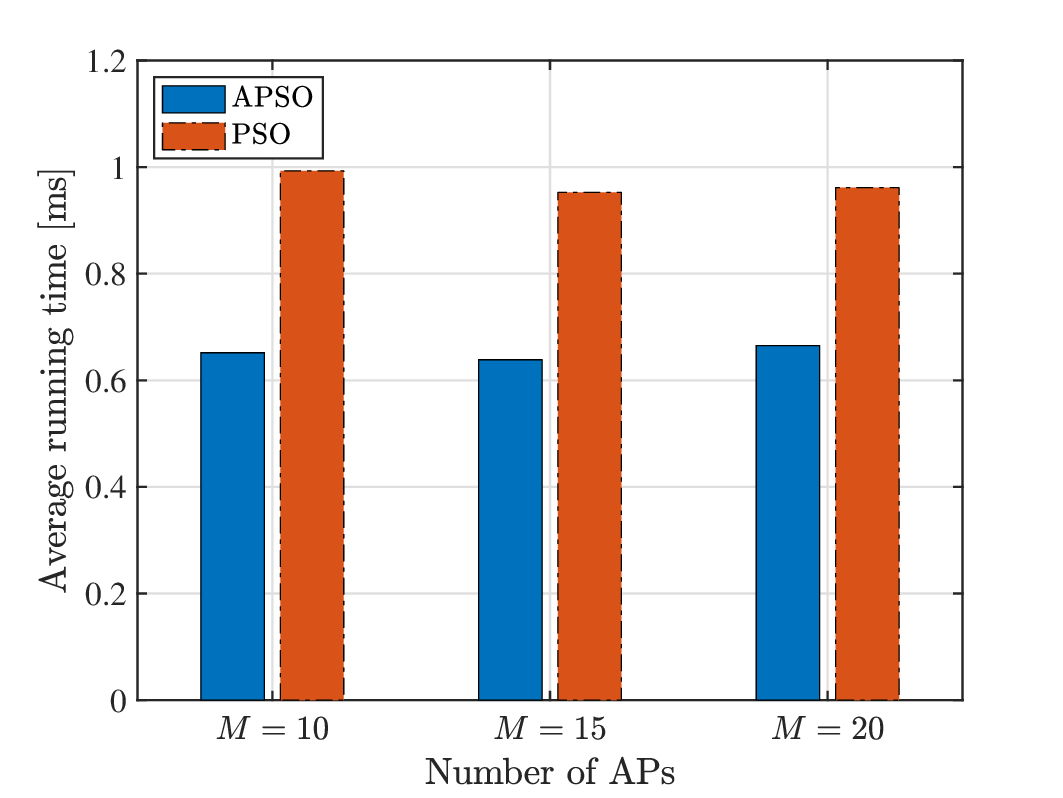}
\caption{\textcolor{black}{Running times of the proposed AO algorithms including APSO and PSO for STAR-RIS-CF-mMIMO with different $M$ ($\gamma_T=\gamma_R=0.8$).}}
\label{Figure11}
\end{figure}

\textcolor{black}{The running times of the AO algorithms using APSO and PSO are depicted in Fig. \ref{Figure11}, where the remaining parameters are the same as in Fig. \ref{Figure10}. An \emph{Inter Core i9-14900HX} processor is employed. It can be observed that given the value of $M$, the APSO-based AO algorithm is more computationally efficient than its PSO-based counterpart, since the AO algorithm with PSO needs more iterations to achieve convergence. In addition, observe that all types of AO algorithms can converge within the coherence interval of $T_c=1$ ms.}

\section{Conclusions}\label{Section 7}
STAR-RIS-CF-mMIMO having realistic imperfect hardware was proposed and its ergodic downlink SE analysis was presented, where both the transceiver hardware impairment and the STAR-RIS phase errors are considered. Then, upon investigating both the pilot contamination and HWIs, the linear MMSE cascade channel estimation was conceived. Moreover, a closed-form expression of the downlink ergodic SE of our systems was derived based on the MR precoding scheme. Our numerical results have shown the accuracy of the SE derived in closed-form by comparing it to our Monte-Carlo simulation results. Furthermore, the joint power control and STAR-RIS passive BF design were investigated to maximize the worst-case ergodic SE among different UEs. To solve the resultant non-convex max-min fairness problems, we decomposed it into two sub-problems, where the proposed APSO and bisection methods were conceived, respectively. The complexity and convergence of the proposed AO-based algorithm were analyzed. Our simulation results illustrated that the STAR-RIS-CF-mMIMO system attains higher SE than its conventional counterparts, while the SE under different hardware parameters was also investigated. Finally, the effectiveness of our AO-based algorithm was validated by its SE enhancements compared to the conventional solutions.
\vspace{-1em}
\begin{appendices}
\section{Useful Results}\label{appendix1}
\emph{Lemma 1 \cite{9973349}:} Let $\pmb{A}\in\mathbb{C}^{L\times N}$ denote a $L\times N$-dimensional matrix whose elements are i.i.d RVs with zero mean and $\xi_a$ variance. Consider a deterministic matrix $\pmb{Z}\in\mathbb{C}^{N\times N}$. Then, we have $\mathbb{E}\left\{\pmb{A}\pmb{Z}\pmb{A}^H\right\}=\xi_a\tr(\pmb{Z})\pmb{I}_L$.

\emph{Lemma 2 \cite{van2021reconfigurable}:} Let us consider a $K$-dimensional vector $\pmb{c}\sim\mathcal{CN}(\pmb{0},\pmb{C})$ with the covariance matrix $\pmb{C}\in\mathbb{C}^{K\times K}$, and a deterministic matrix $\pmb{A}\in\mathbb{C}^{K\times K}$. Then, we can obtain $\mathbb{E}\left\{|\pmb{c}^H\pmb{A}\pmb{c}|^2\right\}=\left|\tr(\pmb{A}\pmb{C})\right|^2+\tr(\pmb{A}\pmb{C}\pmb{A}^H\pmb{C})$.
\vspace{-1em}
\section{Proof of Proposition 1}\label{appendix2}
\subsection{Derivation of $\left|{\tt DS}_k(t)\right|^2$}
Since the channel estimate $\hat{\pmb{h}}_{mk}(0)$ and channel estimation error $\tilde{\pmb{h}}_{mk}(0)$ are uncorrelated, the desired signal term ${\tt DS}_k(t)$ can be formulated as
\begin{align}\label{eq_ap2_DS}
{\tt DS}_k(t)&=\sqrt{\gamma_R\gamma_T\rho}\mathbb{E}\left\{\sum_{m=1}^M\sqrt{\eta_{mk}}\pmb{f}^H_{mk}\hat{\pmb{h}}_{mk}(0)e^{-j\varphi_{mk}(t)}\right\}\nonumber\\
	&\overset{(a)}{=}\sqrt{\gamma_R\gamma_T\rho}\sum_{m=1}^M\sqrt{\eta_{mk}}\tr(\pmb{\Omega}_{mk})e^{-\frac{\delta^2 t}{2}},
\end{align}
where $\chi_{mk}(t)=e^{-j\Delta\varphi_{mk}(t)}$ with $\Delta\varphi_{mk}(t)=\varphi_{mk}(t)-\varphi_{mk}(0)$, while we invoke $\mathbb{E}\left\{e^{j[\varphi_{mk}(t_1)-\varphi_{mk}(t_2)]}\right\}=e^{-\frac{\delta^2|t_1-t_2|}{2}}$ in $(a)$. Here we introduce $\pmb{\eta}_k^{1/2}=\diag\left(\sqrt{\eta_{1k}},\ldots,\sqrt{\eta_{Mk}}\right)\otimes\pmb{I}_L$ and $\pmb{\Omega}_k=\diag\left(\pmb{\Omega}_{1k},\ldots,\pmb{\Omega}_{Mk}\right)$. Hence, we can obtain $\left|{\tt DS}_k(t)\right|^2=\gamma_R\gamma_T\rho e^{-\delta^2 t}\left|\tr(\pmb{\eta}_k^{1/2}\pmb{\Omega}_k)\right|^2$.
\vspace{-1em}
\subsection{Derivation of $\mathbb{E}\left\{|{\tt BU}_k(t)|^2\right\}$}
Let $T_0=\sum_{m=1}^M\sqrt{\eta_{mk}}\pmb{h}^H_{mk}(t)\hat{\pmb{h}}_{mk}(0)$. Then, we have
\begin{align}\label{eq_BU1}
	\mathbb{E}\left\{|{\tt BU}_k(t)|^2\right\}={\gamma_R\gamma_T\rho}\mathbb{E}\left\{\left|T_0\right|^2\right\}-|{\tt DS}_k(t)|^2,
\end{align}
where we can formulate $\mathbb{E}\left\{\left|T_0\right|^2\right\}=T_1+T_2$ with $T_1=\mathbb{E}\left\{\left|\sum_{m=1}^M\sqrt{\eta_{mk}}\hat{\pmb{h}}^H_{mk}(0)\hat{\pmb{h}}_{mk}(0)\chi_{mk}(t)\right|^2\right\}$ and $T_2=\mathbb{E}\left\{\left|\sum_{m=1}^M\sqrt{\eta_{mk}}\tilde{\pmb{h}}^H_{mk}(0)\hat{\pmb{h}}_{mk}(0)\chi_{mk}(t)\right|^2\right\}$. Let  $\hat{\pmb{h}}_k(0)=\left(\hat{\pmb{h}}_{1k}^H(0),\ldots,\hat{\pmb{h}}_{Mk}^H(0)\right)^H$ and $\pmb{\Delta}_k(t)=\diag\left(\chi_{1k}(t),\ldots,\chi_{Mk}(t)\right)\otimes\pmb{I}_L$. Then, $T_1=\mathbb{E}_{\pmb{\Delta}}\left\{\left|\hat{\pmb{h}}_k^H(0)\pmb{\Delta}_k(t)\pmb{\eta}_k^{1/2}\hat{\pmb{h}}_k(0)\right|^2\right\}$ can be reformulated as
\begin{align}\label{eq_BU2}
T_1&=\mathbb{E}_{\pmb{\Delta}}\left\{\left|\tr\left(\pmb{\Delta}_k(t)\pmb{\eta}_k^{1/2}\pmb{\Omega}_k\right)\right|^2\right\}\nonumber\\
&+\mathbb{E}_{\pmb{\Delta}}\left\{\tr\left(\pmb{\Delta}_k(t)\pmb{\eta}_k^{1/2}\pmb{\Omega}_k\pmb{\eta}_k^{1/2}\pmb{\Delta}_k^H(t)\pmb{\Omega}_k\right)\right\}\nonumber\\
	&=\mathbb{E}\left\{\left|\sum_{m=1}^M \tr\left(\sqrt{\eta_{mk}}\pmb{\Omega}_{mk}\chi_{mk}(t)\right)\right|^2\right\}+\tr\left(\pmb{\eta}_k\pmb{\Omega}_{k}^2\right).
\end{align}
Then we can formulate $T_1$ as
\begin{align}\label{eq_BU4}		
T_1&=\sum_{m=1}^M \eta_{mk}|\tr(\pmb{\Omega}_{mk})|^2+\sum_{m=1}^M\eta_{mk}\tr\left(\pmb{\Omega}_{mk}^2\right)\nonumber\\
&+e^{-\varrho^2_{\phi}t }\sum_{m=1}^M\sum_{n\neq m}^M\sqrt{\eta_{mk}\eta_{nk}}\tr(\pmb{\Omega}_{mk})\tr(\pmb{\Omega}_{nk}).\end{align}
Moreover, we have $\mathbb{E}\left\{\tilde{\pmb{h}}_{mk}^H(0)\hat{\pmb{h}}_{mk}(0)\hat{\pmb{h}}_{nk}^H(0)\tilde{\pmb{h}}_{nk}(0)\right\}=0$ with $n\neq m$, since the estimated channels of different APs are independent. Hence, we can derive $T_2=\sum_{m=1}^M\eta_{mk}\tr\left((\pmb{R}_{mk}-\pmb{\Omega}_{mk})\pmb{\Omega}_{mk}\right)$. Therefore, we can obtain $\mathbb{E}\left\{\left|T_0\right|^2\right\}$. By introducing $\pmb{P}_k=\diag(\eta_{1k},\ldots,\eta_{Mk})$, $\pmb{A}_k=\diag(a_{1k},\ldots,a_{Mk})$ with $a_{mk}=|\tr(\pmb{\Omega}_{mk})|^2$ and $\pmb{R}_k=\diag(\pmb{R}_{1k},\ldots,\pmb{R}_{Mk})$, we can finally derive $\mathbb{E}\left\{|{\tt BU}_k(t)|^2\right\}$ as
\begin{align}\label{eq_BU_final}	
&\gamma_T\gamma_R\rho\left\{e^{-\varrho^2_{\phi}t}\left(1-e^{-\varrho^2_{\psi}t}\right)\left|\tr(\pmb{\eta}_k^{1/2}\pmb{\Omega}_k)\right|^2\right.\nonumber\\
&\left.+\left(1-e^{-\varrho^2_{\phi}t}\right)\tr(\pmb{P}_k\pmb{A}_k)+\tr(\pmb{\eta}_k\pmb{R}_k\pmb{\Omega}_{k})\right\}.
\end{align}
\vspace{-2em}
\subsection{Derivation of $\sum_{i\neq k}^{K}\mathbb{E}\left\{|{\tt UI}_{ki}(t)|^2\right\}$}
For the UE interference term, we can write $\sum_{i\neq k}^{K}\mathbb{E}\left\{|{\tt UI}_{ki}(t)|^2\right\}$ as
\begin{align}
\sum_{i\notin\mathcal{P}_k}\mathbb{E}\left\{|{\tt UI}_{ki}(t)|^2\right\}+\sum_{i\in\mathcal{P}_k\setminus\{k\}}\mathbb{E}\left\{|{\tt UI}_{ki}(t)|^2\right\}=T_3+T_4.
\end{align}
Based on the scenarios of $i\notin\mathcal{P}_k$ and $i\in\mathcal{P}_k\setminus\{k\}$, let us denote the non-coherence and coherence interference terms as $T_3=T_{31}+T_{32}$ and $T_4=T_{41}+T_{42}$ according to the conditions of $n=m$ and $n\neq m$, respectively. For the case of $n=m$ and $i\notin\mathcal{P}_k$, we can obtain
\begin{align}\label{eq_UI2}	T_{31}&=\gamma_T\gamma_R\rho\sum_{i\notin\mathcal{P}_k}\sum_{m=1}^M\eta_{mi}\mathbb{E}\left\{\left|\pmb{h}^H_{mk}(0)\hat{\pmb{h}}_{mi}(0)\right|^2\right\}\nonumber\\
&=\gamma_T\gamma_R\rho\sum_{i\notin\mathcal{P}_k}\sum_{m=1}^M\eta_{mi}\tr\left(\pmb{R}_{mk}\pmb{\Omega}_{mi}\right).
\end{align}
Since the channel estimates of different APs are independent,
we can formulate $T_{32}=0$ when $n\neq m$ and $i\notin\mathcal{P}_k$. When $n=m$ and $i\in\mathcal{P}_k\setminus\{k\}$, we have $T_4$ as
\begin{align}
\check{\gamma}\rho\sum_{i\in\mathcal{P}_k\setminus\{k\}}\sum_{m=1}^M\eta_{mi}\mathcal{I}_1+\check{\gamma}\rho\sum_{i\in\mathcal{P}_k\setminus\{k\}}\sum_{m=1}^M\eta_{mi}\mathcal{I}_2,  
\end{align}
where $\check{\gamma}=\gamma_T\gamma_R$, $\mathcal{I}_1=\mathbb{E}\left\{\left|\hat{\pmb{h}}_{mk}^H(0)\hat{\pmb{h}}_{mi}(0)\right|^2\right\}$ and $\mathcal{I}_2=\mathbb{E}\left\{\left|\tilde{\pmb{h}}_{mk}^H(0)\hat{\pmb{h}}_{mi}(0)\right|^2\right\}$.
Since we have $\pmb{z}_{mk}(0)=\pmb{z}_{mi}(0)$ and $\pmb{\Psi}_{mk}=\pmb{\Psi}_{mi}$ under the scenario of $n=m$ and $i\in\mathcal{P}_k\setminus\{k\}$, $\mathcal{I}_1$ can be expressed as
\begin{align}\label{eq_UI5}
	&\gamma_T^2\gamma_R^2 p^2\tau_p^2\mathbb{E}\left\{\left|\pmb{z}^H_{mk}(0)\pmb{\Psi}^{-1}_{mk}\pmb{R}_{mk}\pmb{R}_{mi}\pmb{\Psi}^{-1}_{mi}\pmb{z}_{mi}(0)\right|^2\right\}\nonumber\\
	&\overset{(a)}{=}\gamma_T^2\gamma_R^2 p^2\tau_p^2\mathbb{E}\left\{\left|\pmb{c}^H\pmb{\Psi}^{-1/2}_{mk}\pmb{R}_{mk}\pmb{R}_{mi}\pmb{\Psi}^{-1/2}_{mi}\pmb{c}\right|^2\right\}\nonumber\\
	&\overset{(b)}{=}\gamma_T^2\gamma_R^2 p^2\tau_p^2\left|\tr(\pmb{R}_{mi}\pmb{\Psi}^{-1}_{mk}\pmb{R}_{mk})\right|^2+\tr\left(\pmb{\Omega}_{mk}\pmb{\Omega}_{mi}\right),
		\end{align}
where $\pmb{z}_{mk}(0)\sim\mathcal{CN}(\pmb{0},\pmb{\Psi}_{mk})$ and $\pmb{c}\sim\mathcal{CN}(\pmb{0},\pmb{I}_{L})$ are invoked in $(a)$, while we use Lemma 2 in $(b)$. We can readily show that $\mathcal{I}_2=\tr\left(\left(\pmb{R}_{mk}-\pmb{\Omega}_{mk}\right)\pmb{\Omega}_{mi}\right)$, hence we have $T_{41}=\gamma_T\gamma_R\rho\sum_{i\in\mathcal{P}_k\setminus\{k\}}\sum_{m=1}^M\eta_{mi}\left[\gamma_T^2\gamma_R^2 p^2\tau_p^2\left|\tr(\pmb{R}_{mi}\pmb{\Psi}^{-1}_{mk}\right.\right.\\\left.\left.\pmb{R}_{mk})\right|^2+\tr\left(\pmb{R}_{mk}\pmb{\Omega}_{mi}\right)\right]$. Under the scenario of $n\neq m$ and $i\in\mathcal{P}_k\setminus\{k\}$, we have $T_{42}=\gamma_T\gamma_R\rho e^{-\varrho^2_{\phi}t}\sum_{i\in\mathcal{P}_k\setminus\{k\}}\sum_{m=1}^M\sum_{n\neq m}^M\sqrt{\eta_{mi}\eta_{ni}}\\\mathbb{E}\left\{\hat{\pmb{h}}_{mk}^H(0)\hat{\pmb{h}}_{mi}(0)\right\}\mathbb{E}\left\{\hat{\pmb{h}}_{ni}^H(0)\hat{\pmb{h}}_{nk}(0)\right\}$. Similar to the expression of \eqref{eq_UI5}, we have $\mathbb{E}\left\{\hat{\pmb{h}}_{mk}^H(0)\hat{\pmb{h}}_{mi}(0)\right\}=\gamma_T\gamma_R p\tau_p\tr\left(\pmb{R}_{mi}\pmb{\Psi}_{mk}^{-1}\pmb{R}_{mk}\right)$ and $\mathbb{E}\left\{\hat{\pmb{h}}_{ni}^H(0)\hat{\pmb{h}}_{nk}(0)\right\}=\gamma_T\gamma_R p\tau_p\tr\left(\pmb{R}_{nk}\pmb{\Psi}_{nk}^{-1}\pmb{R}_{ni}\right)$. Consequently we can obtain $T_4=T_{41}+T_{42}$. Let $\pmb{B}_{ik}=\diag(b_{1ik},\ldots,b_{Mik})$ with $b_{mik}=\left|\gamma_T\gamma_R p\tau_p\tr(\pmb{R}_{mi}\pmb{\Psi}_{mk}^{-1}\pmb{R}_{mk})\right|^2$ and $\pmb{\Xi}_{ik}=\diag(\pmb{\Xi}_{1ik},\ldots,\pmb{\Xi}_{Mik})$ with $\pmb{\Xi}_{mik}=\pmb{R}_{mi}\pmb{R}_{mk}^{-1}$ and combine the above derived results, $\sum_{i\neq k}^{K}\mathbb{E}\left\{|{\tt UI}_{ki}(t)|^2\right\}/\gamma_T\gamma_R\rho$ can be obtained as
\begin{align}\label{eq_UI_final}	
&\sum_{i\neq k}\tr(\pmb{\eta}_i\pmb{R}_k\pmb{\Omega}_i)+\sum_{i\in\mathcal{P}_k\setminus\{k\}}e^{-\varrho^2_{\phi}t}\left|\tr\left(\pmb{\eta}_i^{1/2}\pmb{\Xi}_{ik}\pmb{\Omega}_k\right)\right|^2\nonumber\\
&+\left(1-e^{-\varrho^2_{\phi}t}\right)\tr(\pmb{P}_k\pmb{B}_{ik}).
\end{align}
\subsection{Derivation of $\mathbb{E}\left\{\left|{\tt HWI}_k(t)\right|^2\right\}$}
We can express the transmitter HWIs term as $\mathbb{E}\left\{\left|{\tt HWI}_k(t)\right|^2\right\}=\gamma_R\sum_{m=1}^M\mathbb{E}\left\{\left|\pmb{h}^H_{mk}(t)\pmb{\mu}^{\text{AP}}_m(t)\right|^2\right\}=\gamma_R(1-\gamma_T)\rho\sum_{i=1}^K\tr\left(\pmb{\eta}_i\diag(\pmb{\Omega}_i)\pmb{R}_k\right)$.
\subsection{Derivation of $\mathbb{E}\{|{\mu}_{k}^\textit{\rm UE}(t)|^2\}$}
Based on \eqref{eq_re_distortion}, we can formulate $\mathbb{E}\{|{\mu}_{k}^\textit{\rm UE}(t)|^2\}=(1-\gamma_R)\rho\sum_{m=1}^M\sum_{i=1}^K\eta_{mi}\left[\gamma_T\mathcal{I}_3+(1-\gamma_T)\mathcal{I}_4\right]$ with $\mathcal{I}_3=\mathbb{E}\left\{|\pmb{h}_{mk}^H(t)\hat{\pmb{h}}_{mi}(0)|^2\right\}$ and $\mathcal{I}_4=\mathbb{E}\left\{\left \|\hat{\pmb{h}}_{mi}(0)\odot\pmb{h}_{mk}(t)\right \|^2\right\}$. When $i=k$, we have $\mathcal{I}_3=\mathbb{E}\left\{\left|\hat{\pmb{h}}_{mk}^H(t)\hat{\pmb{h}}_{mk}(0)\right|^2\right\}+\mathbb{E}\left\{\left|\tilde{\pmb{h}}_{mk}^H(t)\hat{\pmb{h}}_{mk}(0)\right|^2\right\}$. Moreover, we have $\mathbb{E}\left\{\left|\hat{\pmb{h}}_{mk}^H(t)\hat{\pmb{h}}_{mk}(0)\right|^2\right\}=\mathbb{E}\left\{\left|\hat{\pmb{h}}_{mk}^H(0)\hat{\pmb{h}}_{mk}(0)\right|^2\right\}=\left|\tr(\pmb{\Omega}_{mk})\right|^2+\tr(\pmb{\Omega}_{mk}\pmb{\Omega}_{mk})$ by virtue of Lemma 2, and we can also obtain $\mathbb{E}\left\{\left|\tilde{\pmb{h}}_{mk}^H(t)\hat{\pmb{h}}_{mk}(0)\right|^2\right\}=\mathbb{E}\left\{\left|\tilde{\pmb{h}}_{mk}^H(0)\hat{\pmb{h}}_{mk}(0)\right|^2\right\}=\tr[(\pmb{R}_{mk}-\pmb{\Omega}_{mk})\pmb{\Omega}_{mk}]$. Hence, under the scenario of $i=k$, we can obtain $\mathcal{I}_3=\left|\tr(\pmb{\Omega}_{mk})\right|^2+\tr(\pmb{R}_{mk}\pmb{\Omega}_{mk})$. Based on \eqref{eq_UI2}, we have $\mathcal{I}_3=\tr\left(\pmb{R}_{mk}\pmb{\Omega}_{mi}\right)$ when $i\notin\mathcal{P}_k$. With the help of the results of $\mathcal{I}_1$ and $\mathcal{I}_2$, when $i\in\mathcal{P}_k\setminus\{k\}$ we can obtain $\mathcal{I}_3=\mathcal{I}_1+\mathcal{I}_2=\gamma_T^2\gamma_R^2 p^2\tau_p^2\left|\tr(\pmb{R}_{mi}\pmb{\Psi}^{-1}_{mk}\pmb{R}_{mk})\right|^2+\tr\left(\pmb{R}_{mk}\pmb{\Omega}_{mi}\right)$. Then, we can have $\mathcal{I}_4=\mathbb{E}\left\{\left \|\hat{\pmb{h}}_{mi}(0)\odot\hat{\pmb{h}}_{mk}(t)\right \|^2\right\}+\mathbb{E}\left\{\left\|\hat{\pmb{h}}_{mi}(0)\odot\tilde{\pmb{h}}_{mk}(t)\right \|^2\right\}$. When $i=k$, the first term of $\mathcal{I}_4$ can be derived as $\sum_{l=1}^L\mathbb{E}\left\{\left|\hat{{h}}^{(l)}_{mk}(0)\right|^4\right\}\overset{(a)}{=}2\sum_{l=1}^L\left(\Omega_{mk}^{(l,l)}\right)^2=2\tr\left[\left(\diag(\pmb{\Omega}_{mk})\right)^2\right]$, where $(a)$ is obtained with the help of Lemma 2. Moreover, the second term of $\mathcal{I}_4$ can be calculated as $\sum_{l=1}^L\Omega_{mk}^{(l,l)}C_{mk}^{(l,l)}=\tr[\diag(\pmb{\Omega}_{mk})\diag(\pmb{R}_{mk}-\pmb{\Omega}_{mk})]$. Upon combing the above derived results, we can obtain $\mathcal{I}_4=\tr[\diag(\pmb{\Omega}_{mk})\diag(\pmb{\Gamma}_{mk})]$, where $\pmb{\Gamma}_{mk}=\pmb{R}_{mk}+\pmb{\Omega}_{mk}$. Similarly, when $i\neq k$, we can readily show that $\mathcal{I}_4=\tr[\diag(\pmb{\Omega}_{mi})\diag(\pmb{R}_{mk})]$. Based on the above results of $\mathcal{I}_3$ and $\mathcal{I}_4$ after some algebraic manipulations, we can have $\mathbb{E}\{|{\mu}_{k}^\textit{\rm UE}(t)|^2\}$ as
\begin{align}\label{eq_d4_final}	
&\bar{\gamma}\rho\left[\sum_{i=1}^K\tr(\pmb{\eta}_i\pmb{R}_k\pmb{\Omega}_i)+\tr(\pmb{P}_k\pmb{A}_k)+\sum_{i\in\mathcal{P}_k\setminus\{k\}}\tr(\pmb{P}_k\pmb{B}_{ik})\right]\nonumber\\
+&\tilde{\gamma}\rho\left[\sum_{i=1}^K\tr\left(\pmb{\eta}_i\diag(\pmb{\Omega}_i)\diag(\pmb{R}_k)\right)+\tr\left(\pmb{\eta}_k(\diag(\pmb{\Omega}_{k}))^2\right)\right],
\end{align}
where $\bar{\gamma}=(1-\gamma_R)\gamma_T$ and $\tilde{\gamma}=(1-\gamma_R)(1-\gamma_T)$. Then, by combining the derived results of all the components, we can obtain the final result of $D_k(t)$ as shown in \eqref{eq_Ik}.
\section{\textcolor{black}{Proof of Proposition 2}}\label{appendix3}
\textcolor{black}{Upon defining the variable set as $\mathcal{S}\triangleq\{{\eta}_{mk},\varepsilon_{mk},\varkappa_{ik},q_{mk},r_{mk}\}$, the objective function of \eqref{eq_sub2_p1} can be formulated as
\begin{align}
	f(\mathcal{S})\triangleq\min_{k}\frac{{\gamma_R\gamma_T\rho e^{-\delta^2 t}}\left[\sum_{m=1}^M{z_{mk}}\tr(\pmb{\Omega}_{mk})\right]^2}{\gamma_R\gamma_T\rho e^{-\varrho^2_{\phi}t}\upsilon_{\psi}t\left[\sum_{m=1}^M{z_{mk}}\tr(\pmb{\Omega}_{mk})\right]^2+\mathcal{I}},
\end{align}
where $\upsilon_{\psi}t=1-e^{-\varrho^2_{\psi}t}$, and we have $\mathcal{I}$ shown in \eqref{eq: ap-c-1} at the top of next page. Given $\forall u\in\mathbb{R}^{+}$, the upper-level set associated with $f(\mathcal{S})$ that belongs to $\mathcal{S}$ can be formulated as \eqref{eq_U} which is shown at the top of next page, where we have $\pmb{v}_k\triangleq[\pmb{v}_{k1};\pmb{v}_{k2};\pmb{v}_{k3};\pmb{v}_{k4};\pmb{v}_{k5};\pmb{v}_{k6};1]$ with $\pmb{v}_{k1}=\sqrt{\gamma_T\rho\left(1-\gamma_Re^{-\varrho^2_{\phi}t}\right)}(\pmb{z}_k\odot\pmb{\lambda}_{k1})$, $\pmb{v}_{k2}=\sqrt{\tilde{\gamma}\rho}(\pmb{z}_k\odot\pmb{\lambda}_{k2})$, $\pmb{v}_{k3}=\sqrt{\gamma_T\rho}\pmb{r}_k$, $\pmb{v}_{k4}=\sqrt{\gamma_T\rho\left(1-\gamma_Re^{-\varrho^2_{\phi}t}\right)}\pmb{\varepsilon}_k$, $\pmb{v}_{k5}=\sqrt{\gamma_T\gamma_R\rho e^{-\varrho^2_{\phi}t}}\pmb{\varkappa}_{k}$ and $\pmb{v}_{k6}=\sqrt{(1-\gamma_T)\rho}\pmb{q}_{k}$. Furthermore, we have $\pmb{z}_k=[z_{1k},\ldots,z_{Mk}]^T$, $\pmb{\lambda}_{k1}=\left[\left|\tr(\pmb{\Omega}_{1k})\right|,\ldots,\left|\tr(\pmb{\Omega}_{Mk})\right|\right]^T$, $\pmb{\lambda}_{k2}=[\lambda_{1k2},\ldots,\lambda_{Mk2}]^T$ with $\lambda_{mk2}=\sqrt{\tr\left(\diag(\pmb{\Omega}_{mk})^2\right)}$ for $m=1,\ldots,M$, $\pmb{r}_k=[r_{1k},\ldots,r_{Mk}]$, $\pmb{\varepsilon}_k=[\varepsilon_{1k},\ldots,\varepsilon_{Mk}]^T$, $\pmb{\varkappa}_k=\left\{{\varkappa}_{ik}|{i\in\mathcal{P}_k\setminus\{k\}},\forall i\right\}$ and $\pmb{q}_k=[q_{1k},\ldots,q_{Mk}]^T$. It can be readily shown that the upper-level set $U(f,u)$ is in a second-order cone (SOC) form, hence it is a convex set. Consequently, $f(\mathcal{S})$ and the optimization problem of \eqref{eq_sub2} are all quasi-concave, since the constraint set of \eqref{eq_sub2} is convex.}
\setcounter{eqnback}{\value{equation}} \setcounter{equation}{33}
\begin{figure*}[!t]
\textcolor{black}{\begin{align}\label{eq: ap-c-1}
\mathcal{I}&=\gamma_T\rho\left(1-\gamma_Re^{-\varrho^2_{\phi}t}\right)\sum_{m=1}^M{z_{mk}^2}\left|\tr(\pmb{\Omega}_{mk})\right|^2+\tilde{\gamma}\rho\sum_{m=1}^M{z_{mk}^2}\tr\left(\diag(\pmb{\Omega}_{mk})^2\right)+\gamma_T\rho\sum_{m=1}^Mr^2_{mk}\nonumber\\
	&+\gamma_T\rho\left(1-\gamma_Re^{-\varrho^2_{\phi}t}\right)\sum_{m=1}^M	\varepsilon_{mk}^2+\gamma_T\gamma_R\rho e^{-\varrho^2_{\phi}t}\sum_{i\in\mathcal{P}_k\setminus\{k\}}\varkappa_{ik}^2+(1-\gamma_T)\rho\sum_{m=1}^Mq^2_{mk}+1.
\end{align}}%
\hrulefill
\vspace{-2em}
\end{figure*}
\setcounter{eqncnt}{\value{equation}}
\setcounter{equation}{\value{eqnback}}
\setcounter{eqnback}{\value{equation}} \setcounter{equation}{34}
\begin{figure*}[!t]
\textcolor{black}{\begin{align}\label{eq_U}
U(f,u)&=\{\mathcal{S}:f(\mathcal{S})>u\}=\left\{\mathcal{S}:\frac{{\gamma_R\gamma_T\rho e^{-\delta^2 t}}\left[\sum_{m=1}^M{z_{mk}}\tr(\pmb{\Omega}_{mk})\right]^2}{\gamma_R\gamma_T\rho e^{-\varrho^2_{\phi}t}\left(1-e^{-\varrho^2_{\psi}t}\right)\left[\sum_{m=1}^M\sqrt{\eta_{mk}}\tr(\pmb{\Omega}_{mk})\right]^2+\mathcal{I}}\geq u\right\}\nonumber\\
	&=\left\{\mathcal{S}:||\pmb{v}_k||\leq\sqrt{\left[\left(\frac{1}{u}+1\right)e^{-\varrho_{\psi}^2 t}-1\right]\gamma_R\gamma_T\rho e^{-\varrho^2_{\phi}t}}\sum_{m=1}^M{z_{mk}}\tr(\pmb{\Omega}_{mk})\right\},
\end{align}}
\hrulefill
\vspace{-2em}
\end{figure*}
\setcounter{eqncnt}{\value{equation}}
\setcounter{equation}{\value{eqnback}}
\end{appendices}
\vspace{-1.5em}
	\renewcommand{\refname}{References}
\mbox{} 
\nocite{*}
\bibliographystyle{IEEEtran}
\bibliography{ref.bib}
\end{document}